%% file: main.tex
\documentclass[prl,superscriptaddress,twocolumn,floatfix]{revtex4-1}

\usepackage{amsmath}
\usepackage{amssymb}
\usepackage{bm}         
\usepackage{comment}    
\usepackage{dcolumn}    
\usepackage{graphicx}   
\usepackage[
colorlinks=true,
citecolor=MidnightBlue,
linkcolor=BrickRed,
urlcolor=blue,
breaklinks=true,
]{hyperref}   			
\usepackage{mathtools}	
\usepackage{subfigure}	
\usepackage[
usenames,
dvipsnames
]{xcolor}				

\input{custom-commands}

\graphicspath{{Figs/}}

\makeatletter
\def\@fnsymbol#1{
	\ensuremath{\ifcase#1\or
	*\or				
	\ddagger\or			
	\dagger\or			
	\mathsection\or		
	\mathparagraph\or	
	\|\or				
	**\or				
	\ddagger\ddagger\or	
	\dagger\dagger		
	\else\@ctrerr\fi}}
\makeatother

\begin{document}

%
%

\preprint{APS/123-QED}

\title{Active Contact Forces Drive Non-Equilibrium Fluctuations in Membrane Vesicles}

\author{Sho C. Takatori}
\email{stakatori@ucsb.edu}
\affiliation{
    Department of Chemical Engineering, University of California, Santa Barbara, CA 93106, USA
}

\author{Amaresh Sahu}
\email{amaresh.sahu@berkeley.edu}
\affiliation{
    Department of Chemical \& Biomolecular Engineering, University of California, Berkeley 94720, USA
}

\date{\today}

%
%

\begin{abstract}
	\input{abstract}

\end{abstract}

%
%

\maketitle

%
%

\input{content}

%
%

\input{main.bbl}

%
%

\input{supplemental}

\end{document}

%% file: custom-commands.tex
\newcommand{\kB}{k_{\raisebox{-1pt}{\text{\tiny{B}}}}}
\newcommand{\kBT}{\kB T}
\newcommand{\tauT}{\tau_{\raisebox{-1pt}{\text{\tiny{T}}}}}
\newcommand{\tauP}{\tau_{\raisebox{-1pt}{\text{\tiny{P}}}}}
\newcommand{\tauR}{\tau_{\raisebox{-1pt}{\text{\tiny{R}}}}}

\newcommand{\bmk}{\bm{k}}
\newcommand{\bmx}{\bm{x}}

\newcommand{\bmr}{\bm{r}}
\newcommand{\bmv}{\bm{v}}
\newcommand{\pact}{p^{\mathrm{act}}}
\newcommand{\hatpact}{\hat{p}^{\mathrm{act}}}
\newcommand{\pint}{p^{\mathrm{int}}}
\newcommand{\ptot}{p^{\mathrm{tot}}}
\newcommand{\Nc}{N_{\mathrm{c}}}
\newcommand{\Np}{N_{\mathrm{c}}}
\newcommand{\tsim}{t_{\text{sim}}}

\newcommand{\edit}[1]{\textcolor{blue}{#1}}

%% file: abstract.tex
We analyze the non-equilibrium shape fluctuations of giant unilamellar vesicles encapsulating motile bacteria.
Owing to bacteria--membrane collisions, we experimentally observe a significant increase in the magnitude of membrane fluctuations at low wave numbers, compared to the well-known thermal fluctuation spectrum.
We interrogate these results by numerically simulating membrane height fluctuations via a modified Langevin equation, which includes bacteria--membrane contact forces.
Taking advantage of the length and time scale separation of these contact forces and thermal noise, we further corroborate our results with an approximate theoretical solution to the dynamical membrane equations.
Our theory and simulations demonstrate excellent agreement with non-equilibrium fluctuations observed in experiments.
Moreover, our theory reveals that the fluctuation--dissipation theorem is not broken by the bacteria; rather, membrane fluctuations can be decomposed into thermal and active components.

%% file: content.tex
%
%

\begin{figure}[!b]
	\vspace{-02pt}
	\centering
	\includegraphics[width=1.0\linewidth]{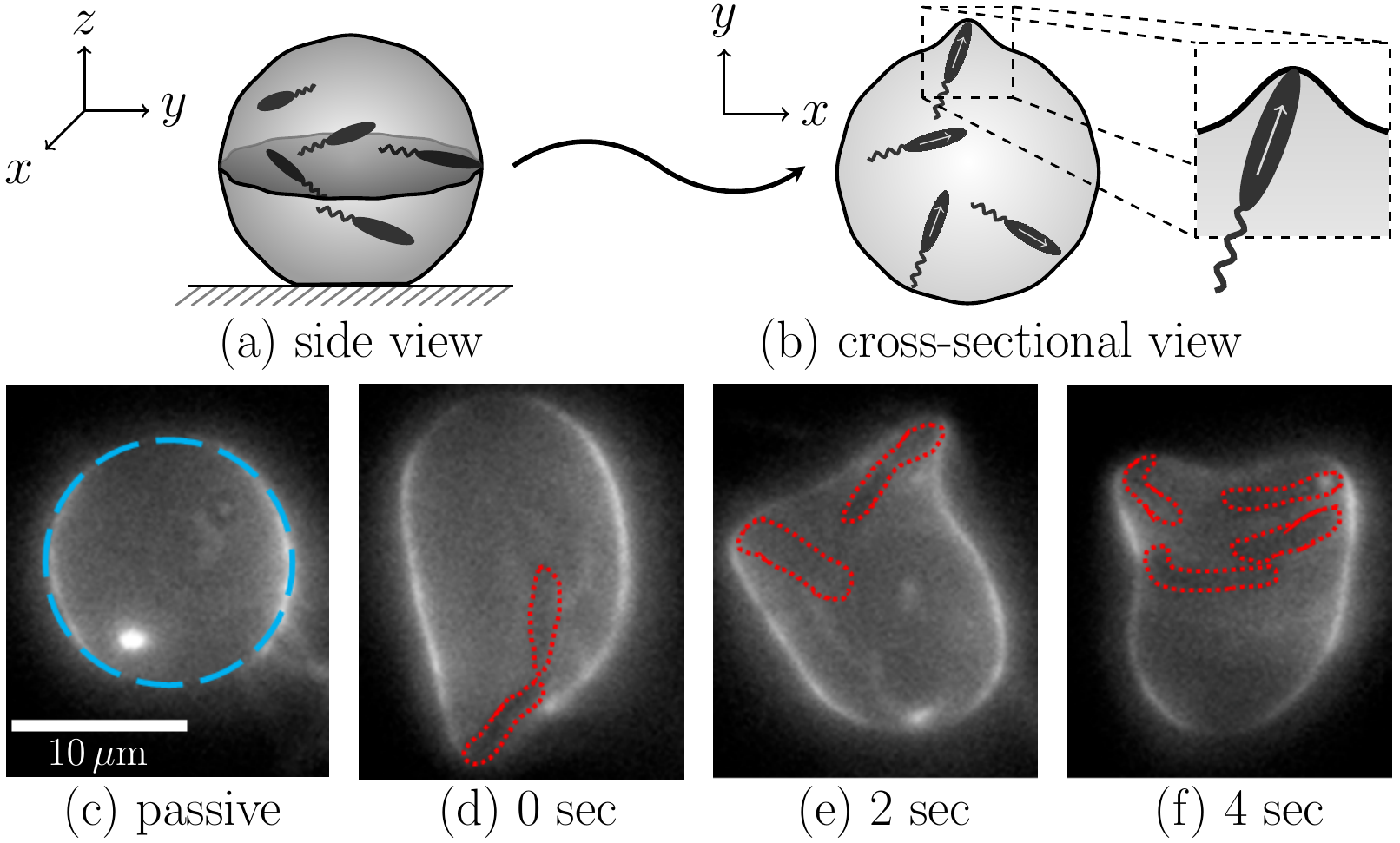}
	\caption{%
		Giant unilamellar vesicle (GUV) containing motile \textit{Bacillus subtilis} PY79.\
		The schematic shows how the three-dimensional system (a) is imaged at a single equatorial cross section (b) to generate the experimental images in (c)--(f).
		The dashed blue outline in (c) shows the undeformed spherical shape of the membrane when bacteria are non-motile, while (d)--(f) show how motile bacteria (dotted red outlines) generate large membrane deformations at different times.
	}
	\label{fig:fig_experiment_small_deformation}
\end{figure}
\begin{figure} [!b]
	\vspace{-10pt}
	\centering
	\includegraphics[width=0.93\linewidth]{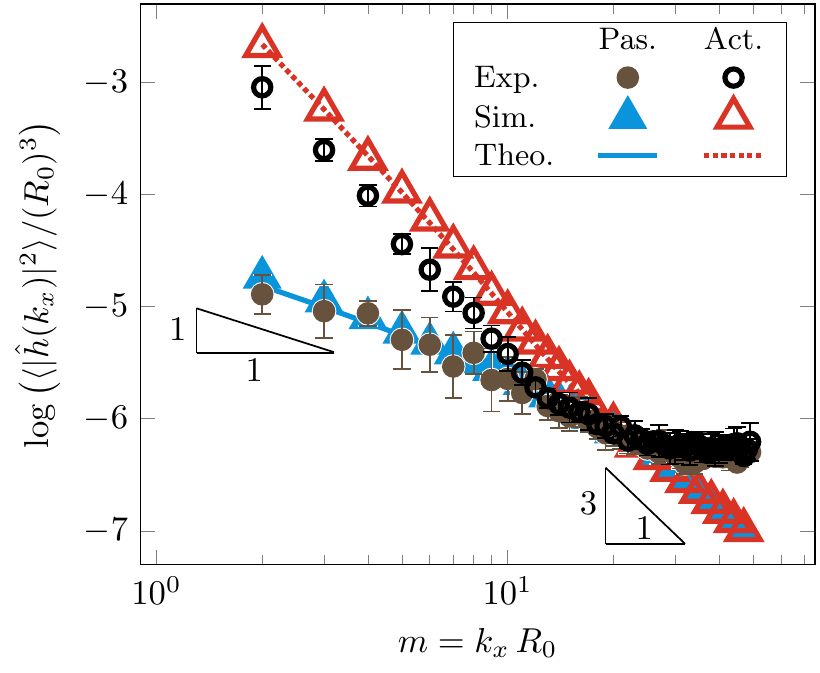}
	\vspace{-10pt}
	\caption{%
		Membrane height fluctuations,
		$\langle |\hat{h}(k_x)|^2 \rangle$,
		for passive (brown, blue) and active (black, red) vesicles, as a function of the mode
		$m = k_x \, R_0$.
		Results are shown from experiments (circles),
		numerical simulations (triangles, see Eq.~\ref{eq:langevin_active_fourier}),
		and analytical theory (lines, see Eq.~\ref{eq:fluctuation_theory}).
		Both simulations and theory show excellent agreement with experiments,
		in the absence of any fitting parameters.
	}
	\label{fig:fig_fluctuation_spectrum}
\end{figure}
Biological lipid membranes make up the boundary of the cell, and act as a dynamic barrier between the cell's internal contents and extracellular environment.
Such membranes are acted upon by a variety of so-called active forces---including those from transmembrane protein pumps \cite{lewis1996directly, balint07} and the underlying cytoskeleton \cite{hackl98, bieling2016force}.
There have been considerable experimental \cite{manneville1999activity, manneville2001active} and theoretical \cite{chen2004internal, gov2004membrane, lomholt2006fluctuation, lin2006nonequilibrium, loubet2012effective, Turlier18, Prost96, ramaswamy2000nonequilibrium, lacoste2005dynamics, ben2011effective, LinBrown04, gov2006dynamics, alert15} efforts to show how active forces from transmembrane proteins and the cytoskeleton cause membrane fluctuations to deviate from the well-known equilibrium result, with a particular emphasis on the membranes of red blood cells \cite{brochard75, Gov03, fournier04, gov2005red, gov2007active, turlier2016equilibrium}.
More recently, there has been growing interest in analyzing the behavior of self-propelled active colloids enclosed within membrane vesicles \cite{Paoluzzi16, Chen17, Chao19, Li19, vutukuri2019sculpting}, as such systems can serve as a useful minimal model of the cell.

In this Letter, we experimentally and theoretically study the membrane shape fluctuations induced by motile bacteria enclosed within giant unilamellar vesicles (GUVs).
A schematic of our experimental system, as well as fluorescence microscopy images involving motile and nonmotile bacteria, are shown in Fig.~\ref{fig:fig_experiment_small_deformation}; see also Vids.\ S1--S5 in the Supplemental Material (SM) \cite{Supplemental}.
We observe motile, micron-sized bacteria pushing against their elastic membrane container and causing large deformations until they reorient after $\sim \! 0.5$ seconds and swim in another direction.
As shown by the filled brown (passive) and open black (active) circles in Fig.~\ref{fig:fig_fluctuation_spectrum}, as well as Fig.\ 1 of the SM \cite{Supplemental}, the bacteria cause a significant change in the distribution of membrane deflections and the corresponding fluctuation spectrum.
Due to the separation in length and time scales of bacteria--membrane contact and equilibrium fluctuations, our active fluctuation spectrum only deviates from its passive counterpart at small wave numbers.
Figure \ref{fig:fig_fluctuation_spectrum} also presents our main quantitative result, as we find excellent agreement between experiments (circles), simulations (triangles), and analytical theory (curves).
We now provide a brief summary of the experimental protocol used to construct the `active vesicles' of Fig.~\ref{fig:fig_experiment_small_deformation} before describing the simulations and analytical theory used to generate Fig.~\ref{fig:fig_fluctuation_spectrum}.

%
%

\textbf{\textit{Experiments.}}---%
A modified electroformation protocol \cite{Angelo75, Kuribayashi06} was used to encapsulate \textit{Bacillus subtilis} PY79 inside GUVs.
A 4 mg/mL stock solution of 99.5\% 1,2-dioleoyl-sn-glycero-3-phosphocholine (DOPC) and 0.5\% L-$\alpha$-phosphatidylethanolamine-N-lissamine rhodamine B sulfonyl (Egg Liss Rhod PE) dissolved in chloroform was spin-coated onto indium tin oxide (ITO) coated glass slides with surface resistivity of $\sim \! 50$--100 $\Omega$/sq.
Luria broth nutrient medium was placed between the ITO slides with a spacer and connected to a wavefunction generator.
After 75--90 minutes of a square wave with 1 V$_{\text{pp}}$ at 10 Hz, a small volume of a dense suspension of an overnight culture of PY79 was added between the ITO slides and set aside in the absence of voltage for 10--15 minutes with the lipid-coated ITO slide facing down.
Finally, we applied 20 minutes of a square wave with 0.3 V$_{\text{pp}}$ at 2 Hz.
The suspension was imaged on an inverted widefield fluorescence microscope at $30^{\circ}$C.

Prior to electroformation, the bacteria are not highly motile, as the overnight culture is in a stationary growth phase.
During electroformation, however, \textit{B.\ subtilis} is introduced into the chamber with fresh nutrient medium; the bacteria become motile after $\sim \! 30$ min
\footnote{Another possibility is that electroformation temporarily weakens the bacteria, and it takes them time to recover.}.
Immediately after electroformation, we identify and image a vesicle containing several nonmotile bacteria to measure the undeformed vesicle radius and the membrane height fluctuations---which correspond to those of a vesicle without bacteria, and which we refer to as a `passive vesicle' (see Fig.~\ref{fig:fig_experiment_small_deformation}c).
Once the bacteria become motile, we measure the membrane fluctuations of the same vesicle
\footnote{Bacterial division occurs on a time scale of $\sim \! 30$--60 min, and so does not affect our measurements}.
In this way, we are able to directly compare passive and active membrane fluctuations of a single vesicle both visually (Fig.~\ref{fig:fig_experiment_small_deformation}c--f and Vids.\ S1--S5 in the SM \cite{Supplemental}) and in Fourier space (Fig.~\ref{fig:fig_fluctuation_spectrum}, filled brown and open black circles).
We analyze the membrane fluctuation spectra of passive and active vesicles using standard methods \cite{Pecreaux04, Gracia10, Meleard11}, in which we have removed the $m = 1$ mode due to experimental difficulties in locating the center of the vesicle
\footnote{We have verified from active particle simulations that small errors in detecting the vesicle center of mass do not significantly affect the results for modes $m \geq 2$.}.
We note that experimental data at large wave numbers level off due to limitations in the camera resolution, whereas our simulations (described subsequently) capture the full spectrum.
Moreover, as we are experimentally capturing fluctuations at only a single cross section of the membrane vesicle (see Fig.~\ref{fig:fig_experiment_small_deformation}), when computing the Fourier spectrum we are implicitly averaging over one of the two independent Fourier modes \cite{Pecreaux04}.

%
%

\textbf{\textit{Development of the theory.}}---%
We have so far experimentally demonstrated how active particles, in this case \textit{B.\ subtilis}, cause dramatic changes to the fluctuation spectrum of the surrounding lipid membrane.
However, the physics underlying such interactions remains unclear.
In particular, while other works have considered active forces arising from transmembrane proteins \cite{Prost96, ramaswamy2000nonequilibrium, chen2004internal, gov2004membrane, lacoste2005dynamics, lomholt2006fluctuation, lin2006nonequilibrium, loubet2012effective, Turlier18, ben2011effective} or simulated active particles in vesicles \cite{Chao19,Li19,Chen17,Paoluzzi16}, there is no theoretical description of our experimental results.
Thus, to better understand our experimental system, we both theoretically and numerically model membrane fluctuations in the presence of active particles.
Both of these developments rely on the so-called Monge parametrization of the membrane \cite{monge}, which treats the membrane as a nearly flat plane with small height perturbations, to avoid the complex equations describing a perturbed spherical membrane \cite{Sahu19}.
Despite this rather severe simplification, the agreement between our experiments, simulations, and theory in the absence of any fitting parameters indicates our simple model captures the essential physics of particle--membrane contact.


In thermal equilibrium, the height fluctuations of a nearly planar membrane described by a Helfrich \cite{Canham1970, Helfrich73, Evans74} Hamiltonian
$
	\mathcal{H}
	= \tfrac{1}{2} \int \!
		\kappa (\nabla^2 h)^2
		+ \lambda (\nabla h)^2
	\, \mathrm{d}x \, \mathrm{d}y
$
are given by
$
	\langle \lvert \hat{h} (\bm{k}) \rvert^2 \rangle_{\mathrm{pas}}
	= \kBT / ( \kappa k^4 + \lambda k^2 )
$,
where
$\bm{k} = (k_x, \, k_y)$
is the wave vector conjugate to position $\bm{x} = (x, \, y)$, $\kBT$ is the thermal energy, $\kappa$ is the membrane bending modulus, and $\lambda$ is the surface tension ($\kappa$ and $\lambda$ are assumed to be constant).
In our experiments, however, the vesicles are only imaged at a single cross section (Fig.~\ref{fig:fig_experiment_small_deformation}).
Thus, to compare experiments and theory, we average the theoretical fluctuation spectrum over $k_y$ modes to find
$
	\langle \lvert \hat{h} (k_x) \rvert^2 \rangle_{\mathrm{pas}}
	= ( k_x^{-1} - (k_x^{\, 2} + \lambda / \kappa)^{-1/2} ) \cdot \kBT / ( 2 \lambda )
$;
details are provided in the SM \cite{Supplemental}.
As shown in Fig.\ \ref{fig:fig_fluctuation_spectrum}, passive experimental data (brown circles) agree with the theoretical prediction, $ \langle \lvert \hat{h} (k_x) \rvert^2 \rangle_{\mathrm{pas}} $, for the choice
$\kappa = 14.3 ~ \kBT$
and
$\lambda = 4 \cdot 10^{-3}$ pN/nm (blue curve).
We fixed these parameters in all of our active membrane calculations, and additionally found our numerical and theoretical active results are insensitive to our choice of $\kappa$ and $\lambda$ \cite{Supplemental}.


Equilibrium techniques cannot describe active vesicle fluctuations due to the presence of non-conservative contact forces, so we turn to a dynamical membrane description.
The Langevin equation governing membrane shape changes is given by \cite{Sapp16, LinBrown04, Turlier18}
\begin{equation} \label{eq:langevin_active}
	\dfrac{\partial h (\bm{x}, t)}{\partial t}
	= \eta (\bm{x}, t)
	+ \int \mathrm{d} \bm{x}'
	\, \Big[
		\Lambda (\bm{x} - \bm{x}') \, 
		p^{\mathrm{tot}} (\bm{x}', t)
	\Big]
	~,
\end{equation}
where $h$ is the membrane height, $\eta$ is Gaussian white noise satisfying the fluctuation--dissipation theorem,
$\Lambda(\bm{x} - \bm{x}') := (8 \pi \mu \lvert \bm{x} - \bm{x}' \rvert)^{-1}$
is the $\bm{e}_z \otimes \bm{e}_z$ component of the Oseen tensor for a Newtonian fluid with viscosity $\mu$, and $-p^{\mathrm{tot}} \bm{e}_z$ is the total force per area exerted on the membrane by the surrounding fluid.
In this case,
$p^{\mathrm{tot}} = p^{\mathrm{int}} + p^{\mathrm{act}}$,
where the internal membrane force per area
$
    p^{\mathrm{int}}
    := - \delta \mathcal{H} / \delta h
    = - \kappa \nabla^4 h + \lambda \nabla^2 h
$,
and $p^{\mathrm{act}}$ is the force per area due to active particles (see SM \cite{Supplemental} for details).

\begin{figure}[!t]
	\begin{center}
		\includegraphics[width=1.0\linewidth]{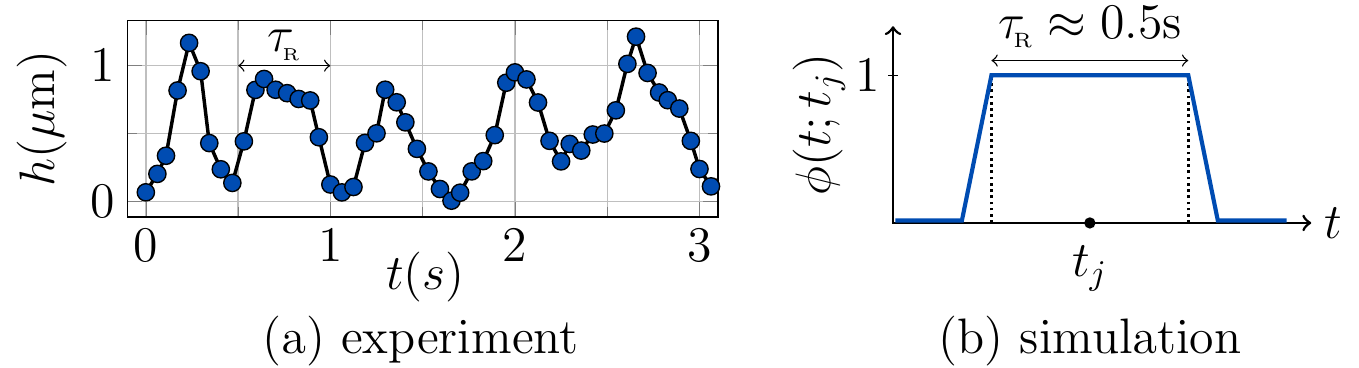}
		\caption{%
			Temporal nature of the bacteria--vesicle collisions.
			(a) Magnitude of the radial deflection of the vesicle at a single location, in a single experiment, as a function of time.
			The trapezoid shape is characteristic of a bacteria pushing against the membrane for reorientation time $\tauR$; smaller peaks indicate a bacteria sliding along the membrane surface.
			(b) Numerical approximation of a head-on collision's temporal component, called $\phi (t; t_j)$.
			The modified step function is centered at the collision time $t_j$.
		}
		\label{fig:fig_temporal_collision}
	\end{center}
	\vspace{-8pt}
\end{figure}

To approximate $p^{\mathrm{act}}$, we model the bacteria as self-propelled particles of half-width $a$ which randomly collide with the membrane vesicle.
For $N_{\mathrm{c}}$ total collisions between the various bacteria and the membrane, where the $j^{\mathrm{th}}$ collision occurs at location $\bm{x}_j$ and time $t_j$, the active force per area on the membrane at location $\bm{x}$ and time $t$ is given by
\begin{equation} \label{eq:active_force}
	p^{\mathrm{act}} (\bm{x}, t)
	= \sum_{j = 1}^{N_{\mathrm{c}}} \,
	\bar{p} \, \phi (t; t_j) \, \exp \Big\{
		- \dfrac{(\bm{x} - \bm{x}_j)^2}{2 a^2}
	\Big\}
	~.
\end{equation}
In Eq.~\eqref{eq:active_force}, $\bar{p}$ is the maximum pressure the bacteria exerts on the membrane, which we estimate to be equal to the pressure exerted by a membrane on a spherical particle of radius $a$,
$\bar{p} \approx 2 \lambda / a$.
Furthermore, as shown in Fig.\ \ref{fig:fig_temporal_collision}, $\phi(t; t_j)$ is a modified step function centered at time $t_j$ which captures the temporal nature of the collision.
In choosing $\phi$, we approximated a bacterium as initially traveling at velocity $U_0$ towards the membrane, coming to rest due to elastic membrane forces, and remaining there for reorientation time $\tauR$ before swimming back into the interior of the vesicle.
Finally, the exponential term in Eq.~\eqref{eq:active_force} is a simple model of the finite size of the particle, which spreads the contact force over a portion of the membrane and is amenable to numerical computation.

At this point, we highlight that all details of the bacteria--membrane interactions are modeled through $\bar{p}$, $\phi (t; t_j)$, and the exponential spreading of the contact force, such that Eq.\ \eqref{eq:active_force} contains the main difference between the present work and other theoretical developments of active membranes \cite{Prost96, ramaswamy2000nonequilibrium, chen2004internal, gov2004membrane, lacoste2005dynamics, lomholt2006fluctuation, lin2006nonequilibrium, loubet2012effective, Turlier18, ben2011effective, LinBrown04, gov2006dynamics, alert15}.
In particular, when active forces arise from membrane--protein interactions, there is no length or time scale separation between active and thermal forces.
As a result, the non-equilibrium fluctuation spectrum can often be obtained by renormalizing the temperature \cite{manneville1999activity, manneville2001active, chen2004internal, gov2004membrane, lomholt2006fluctuation, Turlier18, gov2006dynamics}.
In our case, however, bacteria--membrane interactions are much slower than equilibrium fluctuations, as captured by $\phi$, and are spread over much larger distances, as captured by the Gaussian in Eq.\ \eqref{eq:active_force}.
Note that in our model, for simplicity we neglect the complex hydrodynamic interactions between bacteria and membrane, as well as any permeability effects from fluid passing through the membrane.
As experimental investigations found a rapidly decaying flow field for bacteria close to surfaces \cite{drescher2011fluid}, we simply choose to capture all bacteria--membrane interactions in the active pressure term $p^{\mathrm{act}}$.

%
%

\textbf{\textit{Numerical solution.}}---%
Using standard techniques \cite{Sapp16, Turlier18, LinBrown04}, we take the Fourier transform of Eq.~\eqref{eq:langevin_active} and recognize the Fourier modes are independent.
For each wave vector
$ \bm{k} = (m, n) / R_0$,
where
$m, n \in \mathbb{Z}$
and $R_0$ is the unperturbed vesicle radius,
the corresponding evolution equation is given by \cite{Supplemental}
\begin{equation} \label{eq:langevin_active_fourier}
	\dfrac{\partial \hat{h}(\bm{k}, t)}{\partial t}
	= - \omega(k) \, \hat{h}(\bm{k}, t)
	\, + \, \hat{\eta}(\bm{k}, t)
	\, + \, L \hat{\Lambda} (\bmk) \, \hat{p}^{\mathrm{act}}(\bm{k}, t)
	~.
\end{equation}
In Eq.~\eqref{eq:langevin_active_fourier},
$\omega(k) := (\kappa k^3 + \lambda k)/(4 \mu)$
is the relaxation frequency of mode $\bmk$,
$L = 2 \pi R_0$ is the length of the planar membrane patch,
$\hat{\Lambda} (\bmk) = (4 \mu k L)^{-1}$
is the Fourier transform of $\Lambda (\bmx)$, and $\hat{p}^{\mathrm{act}} (\bm{k}, t)$ is the Fourier transform of the active force per area \eqref{eq:active_force}.
The last term in Eq.~\eqref{eq:langevin_active_fourier} is given by
\begin{equation} \label{eq:active_force_fourier}
	L \hat{\Lambda} (\bmk) \, \hat{p}^{\mathrm{act}}
	= \sum_{j = 1}^{N_{\mathrm{c}}}
	\dfrac{a^2 \, \bar{p}}{4 \mu k R_0} \,
	\phi(t; t_j) \,
	\exp \Big\{
		-i \bm{x}_j \cdot \bm{k}
		- \dfrac{a^2 \, k^2}{2}
	\Big\}
	~.
\end{equation}

We discretize the height evolution equation \eqref{eq:langevin_active_fourier} as shown in the SM \cite{Supplemental} and compute $\hat{h} (\bm{k}, t)$ for all $\bm{k}$, from which we calculate the height fluctuations.
After integrating over $k_y$, we plot our simulation results as the triangles in Fig.~\ref{fig:fig_fluctuation_spectrum} for the passive (filled blue) and active (open red) cases.
Passive results were calculated by setting $\hat{p}^{\mathrm{act}} = 0$ in Eq.~\eqref{eq:langevin_active_fourier}.
While such techniques are known to attain the passive fluctuation spectrum \cite{Sapp16, LinBrown04, Turlier18}, we see excellent agreement between active experiments and simulations as well
\footnote{Our code is publicly available at \texttt{\href{https://github.com/mandadapu-group/active-contact}{https://github.com/ mandadapu-group/active-contact}}}.
Furthermore, there are no fitting parameters in our development:
$\kappa$ and $\lambda$ are found from the membrane fluctuations before bacteria become motile,
the viscosity $\mu$ of the fluid is known,
$R_0 = 4 \, \mu$m is the undeformed vesicle radius,
the bacteria have a reorientation time
$\tauR \approx 0.5$ sec,
and $a = 0.25 \, \mu$m is half the average width of a bacterium.

%
%

\textbf{\textit{Analytical solution.}}---%
To develop an analytical expression for the active membrane fluctuation spectrum, we first consider Eqs.~\eqref{eq:langevin_active_fourier} and \eqref{eq:active_force_fourier} for a vesicle containing a single active particle.
By approximating $\phi(t; t_j)$ as being either 0 or 1 (see Fig.~\ref{fig:fig_temporal_collision}b), the membrane is either fully separated from ($\phi = 0$) or fully in contact with ($\phi = 1$) the bacterium.
When there is no contact, the membrane feels thermal perturbations, such that its height fluctuations are given by the passive result.
If there is contact (denoted with a subscript `c'), the membrane again feels thermal perturbations, but this time oscillates about some nonzero value---which we denote $\bar{h} (\bm{k})$.
In this case, as the time scales of the two processes are separated and the thermal background is independent of the active forces,
the height fluctuations are given by
$
	\langle \lvert \hat{h} (\bm{k}) \rvert^2 \rangle_{\mathrm{c}}
	= \langle \lvert \hat{h} (\bm{k}) \rvert^2 \rangle_{\mathrm{pas}}
	+ \lvert \bar{h} (\bm{k}) \rvert^2
$.
We assume a single bacterium spends reorientation time $\tauR$ in contact with the membrane, then travels for time $\tauT$ in the interior of the vesicle, and repeats.
Thus, for a single particle,
$
	\langle \lvert \hat{h} (\bm{k}) \rvert^2 \rangle
	= \langle \lvert \hat{h} (\bm{k}) \rvert^2 \rangle_{\mathrm{pas}}
	+ \lvert \bar{h} (\bm{k}) \rvert^2 \, \tauR / (\tauR + \tauT)
$.
When there are $N_{\mathrm{p}}$ particles in the vesicle, we assume they are non-interacting, such that the membrane height fluctuations are given by
\begin{equation} \label{eq:active_fluctuation_theory_general}
	\langle \lvert \hat{h} (\bm{k}) \rvert^2 \rangle
	\, = \, \dfrac{\kBT}{\kappa k^4 + \lambda k^2}
	\, + \, \dfrac{N_{\mathrm{p}} \, \tauR}{\tauR + \tauT} \, \lvert \bar{h} (\bm{k}) \rvert^2
	~.
	\vspace{3pt}
\end{equation}
Thus, by determining $\lvert \bar{h} (\bm{k}) \rvert^2$, we determine the membrane fluctuation spectrum of a bacteria-containing lipid membrane vesicle.

To calculate $\bar{h} (\bm{k})$, we average Eq.~\eqref{eq:langevin_active_fourier} in time for the case of a single bacterium, when there is contact ($\phi = 1$).
The time derivative and thermal noise terms average to zero, and $\bar{h} (\bm{k})$ is the average value of $\hat{h} (\bm{k}, t)$.
Thus, by solving for $\bar{h} (\bm{k})$ and substituting into Eq.~\eqref{eq:active_fluctuation_theory_general}, we obtain
\begin{equation} \label{eq:fluctuation_theory}
	\langle \lvert \hat{h} (\bm{k}) \rvert^2 \rangle
	= \, \dfrac{\kBT}{\kappa k^4 + \lambda k^2}
	\, + \, \dfrac{N_{\mathrm{p}} \, \tauR}{\tauT + \tauR}
	\bigg( \!
		\dfrac{a^2 \, \bar{p} / R_0}{\kappa k^4 + \lambda k^2}
	\! \bigg)^{\!\! 2} \, \mathrm{e}^{ - a^2 k^2 }
	\,.
	\vspace{3pt}
\end{equation}
Equation \eqref{eq:fluctuation_theory} is our main theoretical result.
As shown by the dotted red curve in Fig.~\ref{fig:fig_fluctuation_spectrum}, Eq.\ \eqref{eq:fluctuation_theory} demonstrates excellent agreement with the experiments and active simulations---again without any fitting parameters.
Here, the membrane contains
$N_{\mathrm{p}} \approx 7$
bacteria, and we estimate
$\tauT \approx 2 R_0 / U_0 \approx 0.5$ sec
as the time for a bacterium to travel the vesicle diameter, moving at speed
$U_0 \approx 15$ $\mu$m/sec.
We believe our simulations and theory consistently over-predict experimental results because we neglect bacteria--bacteria collisions within the vesicle.
Including such collisions would decrease the number of bacteria--membrane collisions $N_{\text{c}}$ in simulations \eqref{eq:active_force_fourier}, and reduce the proportion of time bacteria are in contact with the membrane in our analytical result \eqref{eq:fluctuation_theory}, both of which would slightly decrease the magnitude of active height fluctuations predicted by theory and simulation.

To test the robustness of our theoretical model, we analyze two additional active vesicles, which are different sizes and contain different numbers of bacteria.
As shown in the SM \cite{Supplemental}, our theory and simulations again demonstrate excellent agreement with experiments when $R_0 \approx 8$ $\mu$m and $\Np \approx 10$, and good agreement when $R_0 \approx 15$ $\mu$m and $\Np \approx 20$.
In the latter case with many bacteria, there are often times when multiple bacteria contact a local portion of the membrane in quick succession---thus violating our assumption of independent bacterial collisions.
Such behavior, which is well-known in the study of active particles near surfaces \cite{yan2015force, nikola2016active}, effectively converts longer wavelength fluctuations into shorter wavelength ones, and qualitatively changes the shape of the active fluctuation spectrum.
We recognize one measure of particle--particle effects at the vesicle boundary is the dimensionless parameter $\Np \, \tauR / (\tauT + \tauR)$ appearing in Eq.\ \eqref{eq:fluctuation_theory}.
In cases where the agreement between experiments and theory is excellent, we calculate
$\Np \, \tauR / (\tauT + \tauR) \approx 3.4$
for the vesicle in Fig.\ \ref{fig:fig_fluctuation_spectrum} and
$\Np \, \tauR / (\tauT + \tauR) \approx 3.2$
for the 10-particle vesicle in the SM \cite{Supplemental}.
In the case where particle--particle correlations become significant at the membrane, however,
$\Np \, \tauR / (\tauT + \tauR) \approx 4.0$%
---which seems to approach the upper limit of our theory's validity.
Thus, we conclude that our theory and simulations are valid in the low-particle regime when
$\Np \, \tauR / (\tauT + \tauR) \lesssim 4$.

%
%

\textbf{\textit{Conclusions.}}---%
Equation \eqref{eq:fluctuation_theory} concludes our theoretical and numerical efforts.
With an analytical expression for the membrane fluctuation spectrum which closely matches experiments, we make several observations regarding the physics of lipid membrane systems driven by active contact forces.
First, Eqs.\ \eqref{eq:active_fluctuation_theory_general} and \eqref{eq:fluctuation_theory} show the fluctuation--dissipation theorem is not broken.
Instead, thermal noise continues to excite all height modes, while active forces dominate small modes.
Intuitively, active contact forces only excite long wavelength modes due to the finite size of a single bacterium, and the distribution of the contact force over a large area.
In fact, the exponential contact term in Eq.\ \eqref{eq:fluctuation_theory} is the main difference between the present work and those concerned with active fluctuations of transmembrane proteins \cite{gov2004membrane, ben2011effective, lin2006nonequilibrium}: by setting the protein timescale to be large in the latter, one recovers an expression similar to Eq.\ \eqref{eq:fluctuation_theory}---however without the exponential term.
Additionally, our analytical result \eqref{eq:fluctuation_theory} demonstrates the active fluctuation spectrum does not follow a power-law behavior at low $\bm{k}$, and for this reason we do not provide a scaling relation in the active region of Fig.\ \ref{fig:fig_fluctuation_spectrum}.
Importantly, our theory and simulations took advantage of the time and length scale separation between active contact and equilibrium forces, and as a result we were able to capture the essential membrane physics using simple techniques.
We note that our theoretical prediction is robust to variations in bacterial and membrane properties, as demonstrated by our sensitivity analysis in the SM \cite{Supplemental}.

\begin{figure}[!t]
	\begin{center}
		\includegraphics[width=0.97\linewidth]{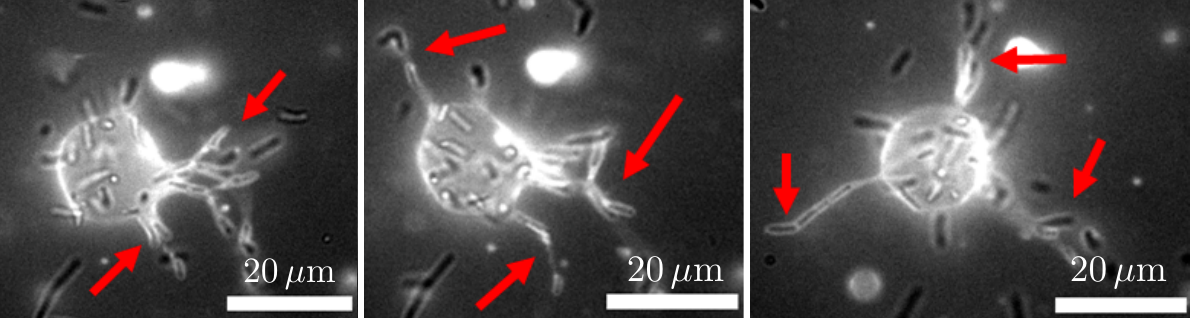}
		\caption{%
			Experimental images of motile \textit{B.\ subtilis} contained within a GUV with low bending modulus $\kappa$ and surface tension $\lambda$.
			When the vesicle is soft, the bacteria generate long membrane tubes upon collision (red arrows).
			Other than membrane bending stiffness and surface tension, experimental conditions are identical to those of Fig.~\ref{fig:fig_experiment_small_deformation}.
		}
		\label{fig:fig_experiment_large_deformation}
	\end{center}
	\vspace{-18pt}
\end{figure}

We end this Letter by providing two avenues for future directions.
First, our experimental method can be easily adapted to encapsulate different types of active particles.
As one example, we synthesized active Janus particles as in Ref.~\cite{Takatori16a}, encapsulated them in lipid membrane vesicles using similar experimental methods, and induced them to propel with 0.5--2.0\% hydrogen peroxide (see Vids. S6 and S7 in the SM \cite{Supplemental}).
Janus particles may also be synthesized with a thin layer of ferromagnetic material embedded underneath the final catalytic layer \cite{Baraban13}, such that by encapsulating them in a vesicle, one would obtain a fully synthetic, stimuli-responsive lipid membrane vesicle.

In addition to changing the active constituents of a membrane vesicle, one could also investigate vesicles with different membrane properties.
In particular, electroformation results in vesicles with a wide range of physical parameters, from which vesicles with specific properties can be selected.
Figure \ref{fig:fig_experiment_large_deformation}, for example, shows a vesicle with low bending modulus $\kappa$ and surface tension $\lambda$ which contains $\approx \! 12$ motile \textit{B.\ subtilis} bacteria (see Vid.\ S8 in the SM \cite{Supplemental}).
For this set of material parameters, the elastic membrane restoring force cannot balance propulsive bacterial forces, and the bacteria form long, protruding tubes.
These membrane tubes, which can be tens of microns in length, persist until the bacteria reorient and swim back towards the vesicle center.
Bacteria--membrane systems such as those shown in Fig.\ \ref{fig:fig_experiment_large_deformation} may be useful as a synthetic model of an infected mammalian cell: several human pathogens, including \textit{Listeria} and \textit{Shigella}, are known to undergo actin-based motility, deform the cell membrane to form membrane tubes, and tunnel into neighboring host cells \cite{Friedrich12, Pizarro16}.
Large membrane shape changes beyond the linear limit have also recently been observed in simulations and experiments \cite{vutukuri2019sculpting, fosnaric2019theoretical, Li19}; in some cases where $\Np \, \tauR / (\tauT + \tauR)$ was large, a spherical-to-prolate vesicle shape change was observed.
To model such highly nonlinear deformations, the full membrane equations \cite{Sahu17} and advanced numerical methods \cite{Sahu18} are required.

%
%

\smallskip
\textbf{\textit{Acknowledgements.}}---%
S.C.T.\ would like to thank John Brady for valuable support and discussions, Heun-Jin Lee, Rob Phillips, and Mikhail Shapiro for integral support with experiments, and Griffin Chure for generous donation of \textit{B.\ subtilis} PY79. 
A.S.\ would like to thank Kranthi Mandadapu for many stimulating discussions, as well as David Limmer for his feedback on the initial simulations, which were submitted as part of a graduate course at U.C.\ Berkeley.
\\[2pt]
\indent S.C.T.\ acknowledges support from the Miller Institute for Basic Research in Science at U.C.\ Berkeley.
A.S.\ is supported by the Computational Science Graduate Fellowship from the U.S.\ Department of Energy, as well as U.C.\ Berkeley.

%% file: main.bbl
%

%% file: supplemental.tex
%
%

\onecolumngrid

\def\thesection{\arabic{section}}
\def\thesubsection{\arabic{section}.\arabic{subsection}}
\def\thesubsubsection{\arabic{section}.\arabic{subsection}.\arabic{subsubsection}}

\onecolumngrid

%
%

\newpage
\clearpage

\begin{center}
	\textbf{\large Active Contact Forces Drive Non-Equilibrium Fluctuations in Membrane Vesicles} \\[7pt]
	\textbf{\large Supplemental Material} \\[16pt]
	Sho C.\ Takatori and Amaresh Sahu
\end{center}

%
%

\input{sm-content}

%
%

\input{sm.bbl}

%% file: sm-content.tex
%
%

\section{1. Experimental Methodology} \label{sec:sec_experimental_methodology}

The main text contains a description of our experimental methods; in this section, we provide additional experimental details.
Membrane fluctuations were measured by epifluorescence microscopy using a Nikon Eclipse Ti inverted microscope with a 60x/NA 1.4 Plan Apo objective.
We recorded hundreds of consecutive images of the equatorial cross-section of a vesicle with a digital CCD camera, with an exposure time of 50 ms.
An in-house code, based on Canny edge detection, was used to detect the edges of the membrane vesicle, and existing methods were applied to compute the transverse height fluctuations of giant unilamellar vesicles \cite{sm-Pecreaux04, sm-Tsai11}.

The positions of the membrane edge are projected onto a Fourier series with 50 modes, according to
\begin{equation} \label{eq:expFT}
r(\theta,t) = R(t) \left( 1 + \sum_{m=1}^{50} a_m \cos(m\theta) + b_m \sin(m\theta) \right),
\end{equation}
where $R(t)$ is the vesicle radius at time $t$ and $m$ is the mode number.
The height fluctuations of the membrane are given by
\begin{equation} \label{eq:expHeight}
\big \langle
		\big \lvert \hat{h} (k_x, t) \big \rvert^2
	\big \rangle
	= \frac{\pi R_0^3}{2} \left( \langle \lvert c_m \rvert^2 \rangle - \langle \lvert c_m \rvert \rangle^2 \right),
\end{equation}
where $R_0 = \langle R(t) \rangle$ is the time-averaged vesicle radius, $k_x = m/R_0$ is the wave vector, and the Fourier coefficients $\lvert c_m \rvert = ( a_m^2 + b_m^2 )^{1/2}$.
As only the transverse fluctuations along the equatorial cross-section of the vesicle are captured in the experiments, our data is implicitly averaged over longitudinal, out-of-focus fluctuations.
Accordingly, we average our analytical theory over one of the two independent modes, such that our passive experimental results can be compared to equilibrium theory.

In practice, one long experimental acquisition was broken into 30 independent segments, and the fluctuations were computed for each segment.
All experimental results in this work report a mean over these independent segments, with the relative error computed as $0.434 \times  \sigma(\langle \lvert \hat{h}\rvert^2 \rangle) / \chi(\langle \lvert \hat{h} \rvert^2 \rangle$---where $\sigma(z)$ and $\chi(z)$ are the standard deviation and mean of a set of data $z$.
We use the method described in Ref.\ \cite{sm-Baird94} to report symmetric error bars on a logarithmic scale.

As noted in other studies \cite{sm-Meleard92, sm-Meleard11}, fluctuations with a lifetime shorter than the integration time of the camera (i.e.\ aperture time of the camera shutter) are not correctly fitted.
For the active vesicles, where fluctuation amplitudes are large and long lasting, we do not anticipate the finite camera integration time to influence our results.

\begin{figure}[p]
	\centering
	\subfigure[\ Instantaneous snapshot of a giant unilamellar vesicle containing motile bacteria.]{\includegraphics[width=0.37\linewidth]{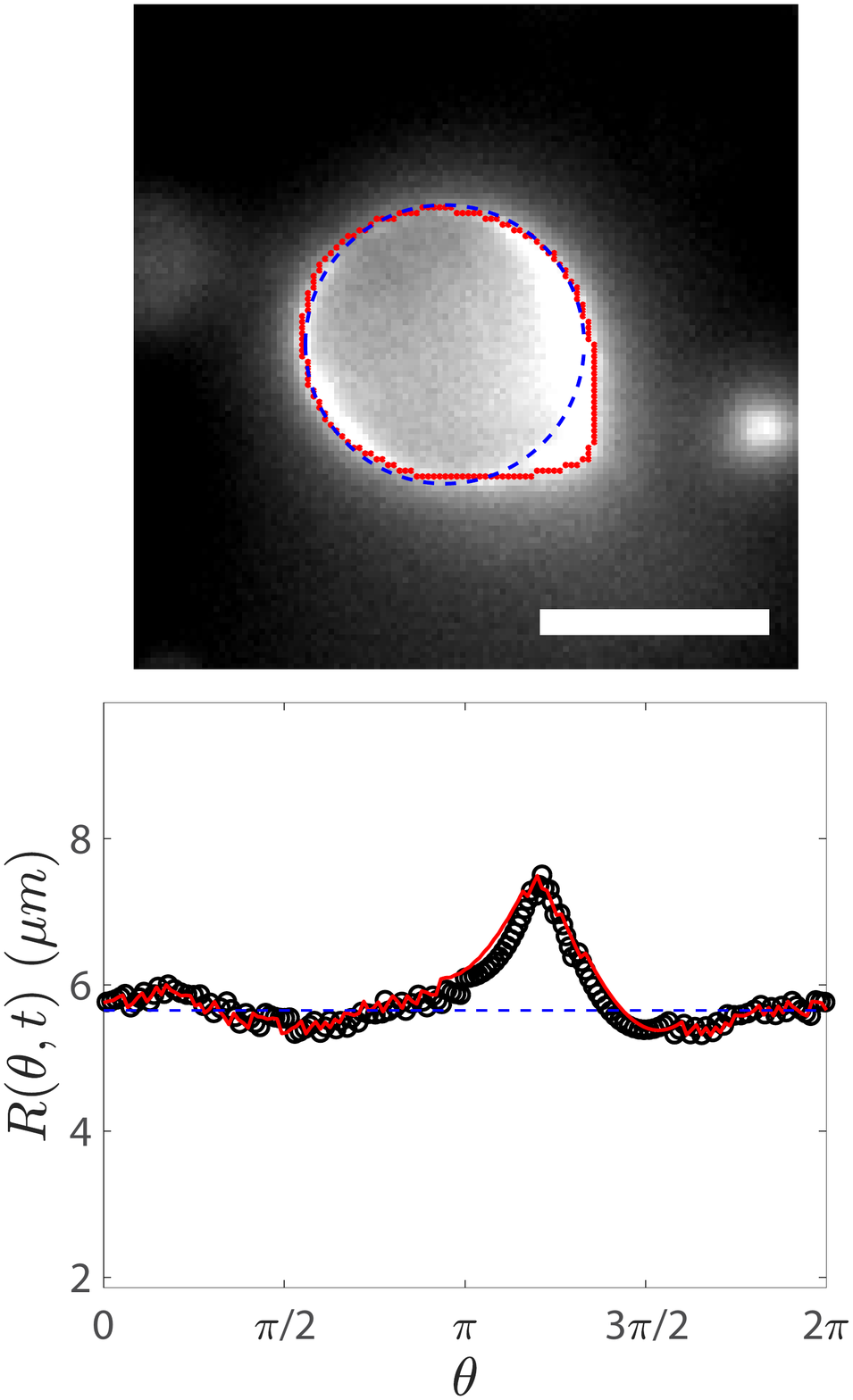}
	\label{fig:fig_edge_detection}}
	\subfigure[\ Distribution of membrane deflections with and without active forces]{\includegraphics[width=0.37\linewidth]{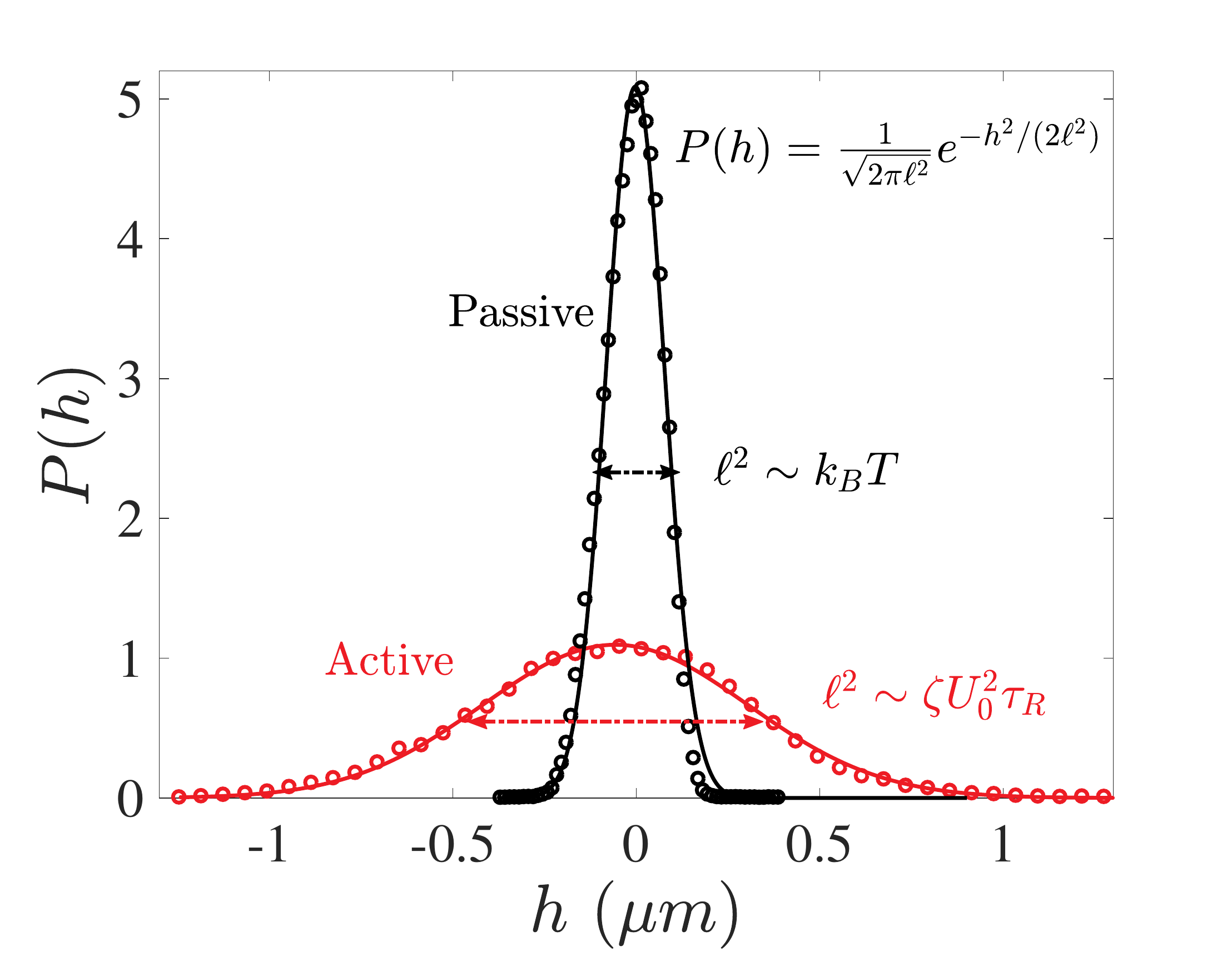}
	\label{fig:fig_probability_deflection}}
	\caption{(a) Instantaneous fluorescence image of a giant unilamellar vesicle containing motile \textit{B. subtilis} (not visible), and corresponding Fourier transform analysis.
	Above, blue dashed circle corresponds to the vesicle baseline position about its center, and the red dots indicate the location of the vesicle membrane edge.
	The scale bar is 10 $\mu$m.
	Below, black circles are the radial positions along the vesicle edge, and the red curve is the Fourier series to the data.
	(b) Normalized probability distribution of membrane deflections about the mean vesicle radius, for passive (black symbols) and active (red symbols) vesicles.
	The distribution was computed by binning over the angular positions around the vesicle and measuring the height deflection from the radial profile from (a).
	Solid curves are a fit to a Gaussian distribution, where $\ell$ is the width of the distribution.}
	\label{fig:fig_experimental_probability}
\end{figure}
\begin{figure}[p]
	\centering
	\subfigure[\ Passive membrane. ]{\includegraphics[width=0.37\linewidth]{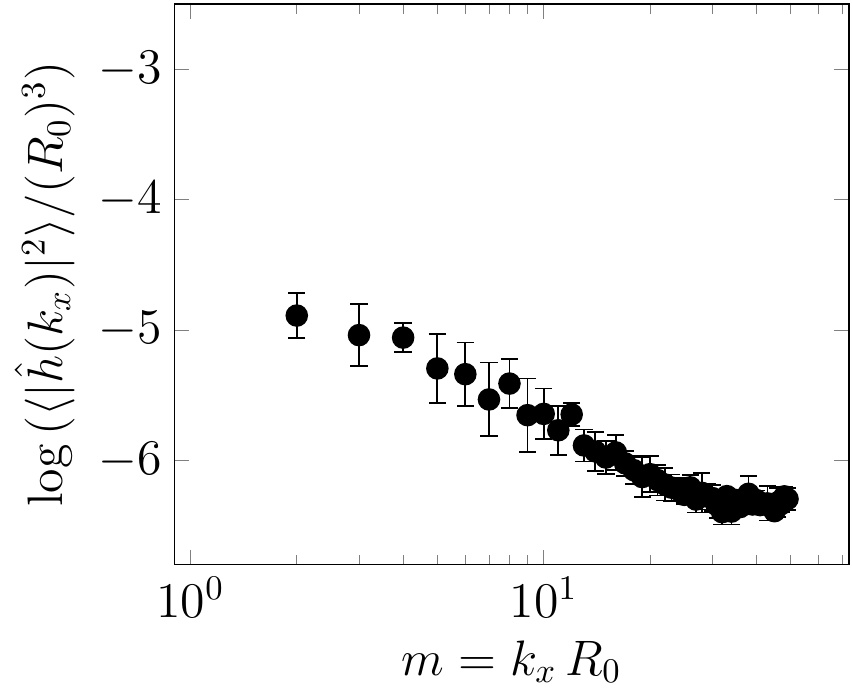}
	\label{fig:fig_experiments_passive}}
	\hspace{20pt}
	\subfigure[\ Active membrane.
	]{\includegraphics[width=0.37\linewidth]{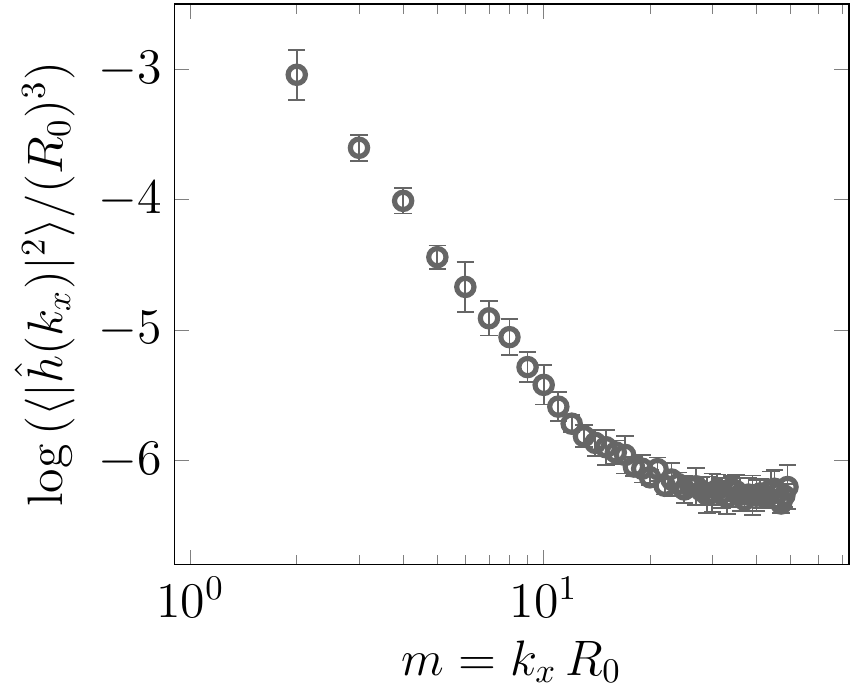}
	\label{fig:fig_experiments_active}}
	\caption{Membrane shape fluctuation spectra of giant unilammelar vesicles containing several non-motile (left) and motile (right) \textit{B. subtilis} PY79.
	Height fluctuations $\big \langle \big \lvert \hat{h} (k_x) \big \rvert^2 \big \rangle$ are nondimensionalized with the average vesicle radius $R_0$, and plotted as a function of the mode number $m = k_x R_0$.
	The data above are plotted together in Fig.\ 2 of the main text.
	Error bars are reported as described in Sec.\ 1, and include measurements from 30 independent time trajectories on the same vesicle.}
	\label{fig:fig_experiments}
\end{figure}

%
%

\subsection{1.1 Results} \label{sec:sec_experimental_results}
Here, we present experimental results, using the methodology described above to compute the Fourier transform of vesicle deformations as well as their fluctuation spectrum.
Figure \ref{fig:fig_edge_detection} shows an instantaneous snapshot of a vesicle with a protrusion caused by contact forces of a motile \textit{B. subtilis} (top), and the corresponding radial profile of the vesicle edge about its center (bottom).
Figure \ref{fig:fig_probability_deflection} is the probability distribution of membrane deflections experienced by the vesicle containing non-motile (`passive', in black symbols) and motile (`active', in red symbols) bacteria.
Solid curves are Gaussian distributions, where the width $\ell$ is a function of membrane bending stiffness, tension, and the relevant driving force of the fluctuations.
For passive vesicles, $\ell$ is governed by the thermal energy $k_B T$, whereas the active vesicles have a distribution governed by the activity scale $\zeta U_0^2 \tau_R$, where $\zeta$ is the hydrodynamic drag factor on the motile bacteria, $U_0$ is the swimming speed, and $\tau_R$ is the reorientation time of the bacteria.
Because the activity scale $\zeta U_0^2 \tau_R \gg k_B T$, the active probability distribution is significantly wider than its passive counterpart, as shown in Fig.\ \ref{fig:fig_probability_deflection}.

The aforementioned probability distributions demonstrate that when vesicles contain motile bacteria, the magnitude of membrane deformation increases.
We infer further information about the membrane deflections by plotting the height fluctuation spectra, which are calculated according to Eqs.\ \eqref{eq:expFT} and \eqref{eq:expHeight}.
Figure \ref{fig:fig_experiments} shows the fluctuation spectrum for passive (a) and active (b) vesicles.
Comparing the two cases, there is a significant increase in magnitude of the fluctuations, however only at low modes.
In the subsequent sections, we derive a theory that elucidates the underlying physics of these active fluctuations.

%
%

\section{2. Theory and Simulation of Passive Membranes} \label{sec:sec_passive_membranes}

In this section, we model lipid membrane vesicles in thermal equilibrium with the surrounding fluid, following well-established techniques \cite{sm-Sapp16, sm-LinBrown04}.
First, equilibrium statistical mechanics is used to determine the membrane fluctuation spectrum.
As equilibrium methods cannot be used to study the active membrane system of interest, we next present a dynamical equation involving membrane--fluid interactions, which is shown to recover the same fluctuation spectrum.
Finally, we describe our methodology to simulate lipid membrane dynamics, which again is amenable to the addition of active forces, and provide our numerical results.
We note that none of the theoretical or computational results in this section are new.
Rather, we present these results for clarity, prior to extending them to active systems in subsequent sections.

%
%

\subsection{2.1. Equilibrium Theory} \label{sec:sec_passive_equilibrium_theory}

We begin by considering a fluctuating lipid membrane in thermal equilibrium at temperature $T$.
The Hamiltonian $\mathcal{H}$ of such a system was determined in the seminal works of P.\ B.\ Canham \cite{sm-Canham1970}, W.\ Helfrich \cite{sm-Helfrich73}, and E.\ A.\ Evans \cite{sm-Evans74}, and was found to be given by
\begin{equation} \label{eq:helfrich_hamiltonian}
	\mathcal{H}
	= \int \Big(
		2 \kappa H^2
		+ \lambda
	\Big)
	\, \mathrm{d}a
	~.
\end{equation}
In Eq.\ \eqref{eq:helfrich_hamiltonian}, $\kappa$ is the elastic bending modulus, $H$ is the mean curvature, $\lambda$ is the surface tension, and the integral is over the membrane surface.
The first term in the integral in Eq.~\eqref{eq:helfrich_hamiltonian} accounts for the energetic cost of membrane bending, while the second term describes the energetic cost of creating additional area.

\begin{figure}[!b]
	\begin{center}
		\includegraphics[width=0.40\linewidth]{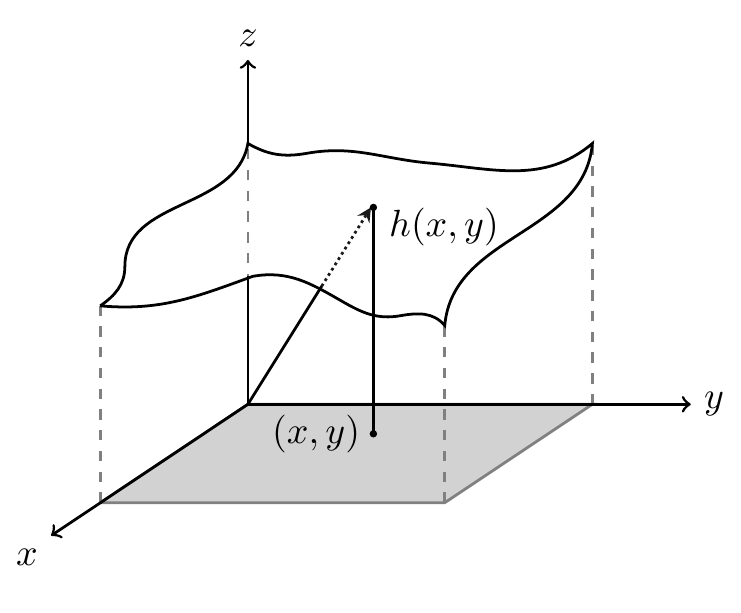}
		\caption{
			A nearly planar lipid membrane patch.
			The membrane height $h(x, y)$ is specified above every point $(x, y)$ in the $x$--$y$ plane.
			The gray region depicts the $[0, L] \times [0, L]$ square over which the membrane is modeled, with periodic boundary conditions.
		}
		\label{fig:fig_monge}
	\end{center}
\end{figure}
While lipid membranes may in general undergo arbitrarily large deformations, the present study is limited to modeling the simpler case of nearly planar membranes undergoing only small out-of-plane deformations.
To describe such a membrane, the membrane height $h(\bmx, t)$ is specified above every point
$\bmx = (x, y)$
in the $x$--$y$ plane (Fig.~\ref{fig:fig_monge}).
The aforementioned surface description is called a Monge parametrization \cite{sm-monge}, and is commonly used in the description of nearly planar membrane systems.
A membrane patch with periodic boundary conditions is considered, such that the region associated with one period lies above an $L \times L$ square in the $x$--$y$ plane.
For the case of small deformations, only terms up to second order in the height $h$ are kept in the Hamiltonian \eqref{eq:helfrich_hamiltonian}, which simplifies to
\begin{equation} \label{eq:helfrich_hamiltonian_flat}
	\mathcal{H}
	= \dfrac{1}{2} \int \Big(
		\kappa (\nabla^2 h)^2
		+ \lambda (\nabla h)^2
	\Big)
	\, \mathrm{d}a
	~.
\end{equation}

As described in the main text, it is sometimes useful to describe lipid membrane fluctuations in Fourier space.
To this end, the two-dimensional Fourier transform and inverse Fourier transform are respectively defined as
\begin{align}
	\hat{h} (\bmk, t)
	&= \dfrac{1}{L} \int \mathrm{d}\bmx ~ \mathrm{e}^{-i \bmk \cdot \bmx} \, h(\bmx, t)
	\label{eq:fourier}
	\\
	\shortintertext{and}
	h (\bmx, t)
	&= \dfrac{1}{L} \sum_{\bmk} \hat{h} (\bmk, t) \, \mathrm{e}^{i \bmk \cdot \bmx}
	~.
	\label{eq:fourier_inverse}
\end{align}
The inverse Fourier transform \eqref{eq:fourier_inverse} sums only over discrete wave vectors $\bmk$ due to the periodic boundary condition requirement.
By substituting Eq.~\eqref{eq:fourier_inverse} into Eq.~\eqref{eq:helfrich_hamiltonian_flat}, and assuming different bending modes are independent, one obtains
\begin{equation} \label{eq:helfrich_hamiltonian_flat_fourier}
	\mathcal{H}
	= \dfrac{1}{2} \sum_{\bmk} \Big(
		\kappa k^4
		+ \lambda k^2
	\Big) \, \big \lvert \hat{h} (\bmk) \big \rvert^2
	~.
\end{equation}
Applying the equipartition theorem to Eq.~\eqref{eq:helfrich_hamiltonian_flat_fourier}, the passive membrane fluctuation spectrum is found to be
\begin{equation} \label{eq:helfrich_fluctuation_spectrum}
	\big \langle
		\big \lvert \hat{h} (\bmk) \big \rvert^2
	\big \rangle_{\text{pas}}
	= \dfrac{\kB T}{\kappa k^4 + \lambda k^2}
	~.
\end{equation}

To compare experimental measurements of lipid membrane fluctuations to theoretical results, we recognize experimental images are captured only at a single cross-section of the vesicle (see Fig.\ 1(a) in the main text).
Thus, to compare with experimental results, the membrane fluctuation spectrum is averaged over all $k_y$ modes according to
\begin{align}
	\big \langle
		\big \lvert \hat{h} (k_x) \big \rvert^2
	\big \rangle
	&:= \dfrac{1}{2\pi} \int_{- \infty}^\infty
		\big \langle
			\big \lvert \hat{h} (\bmk, t) \big \rvert^2
		\big \rangle
	~\mathrm{d} k_y
	~.
	\label{eq:fluctuation_averaging}
	\\
	\intertext{In the case of a passive vesicle in thermal equilibrium with the surrounding fluid, we substitute Eq.~\eqref{eq:helfrich_fluctuation_spectrum} into Eq.\ \eqref{eq:fluctuation_averaging} to obtain}
	\big \langle
		\big \lvert \hat{h} (k_x) \big \rvert^2
	\big \rangle_{\text{pas}}
	&= \dfrac{\kB T}{2 \lambda} \bigg(
		\dfrac{1}{k_x}
		- \dfrac{1}{\sqrt{k_x^{\, 2} + \lambda / \kappa}}
	\bigg)
	~.
	\label{eq:helfrich_fluctuation_spectrum_averaged}
\end{align}
Equation \eqref{eq:helfrich_fluctuation_spectrum_averaged} is used to compare theoretical and experimental results, and is plotted in Fig.~\ref{fig:fig_results_passive} as well as Fig.~2 of the main text.

%
%

\subsection{2.2. Non-Equilibrium Theory} \label{sec:sec_passive_nonequilibrium_theory}

The equilibrium results presented thus far rely on the equipartition theorem, which is not applicable in the presence of active forces.
Consequently, in this section we describe a non-equilibrium theory which
(i) models a lipid membrane sheet fluctuating in a Newtonian fluid,
(ii) reproduces the membrane fluctuation spectrum \eqref{eq:helfrich_fluctuation_spectrum}, and
(iii) is amenable to modeling active forces.
We first describe the general continuum equation describing the lipid membrane shape, and then show how effects from the solvent are included.
While the results of this section are well-known \cite{sm-Sapp16, sm-LinBrown04}, we introduce ideas such that they can be easily extended to the case of active membranes.

%
%

\subsubsection{2.2.1. General Dynamical Equation of a Lipid Membrane} \label{sec:sec_dynamical_equation_without_fluid}

For a nearly planar membrane without a base flow, the linearized equation governing the membrane shape is given by
\begin{equation} \label{eq:shape_equation_no_fluid}
	0
	= [p]
	+ \lambda \nabla^2 h
	- \kappa \nabla^4 h
	~,
\end{equation}
where $[p]$ is the jump in the normal traction across the membrane surface.
The two other terms in Eq.~\eqref{eq:shape_equation_no_fluid} describe the internal membrane forces, arising from surface tension and bending effects, respectively, and have units of pressure.
For notational convenience, we define the internal membrane force per area $\pint$ as
\begin{equation} \label{eq:internal_membrane_pressure}
	\pint
	:= \lambda \nabla^2 h
	- \kappa \nabla^4 h
	~,
\end{equation}
such that Eq.~\eqref{eq:shape_equation_no_fluid} can be written as
$0 = [p] + \pint$.

%
%

\subsubsection{2.2.2. Dynamical Equation with Surrounding Fluid} \label{sec:sec_dynamical_equation_with_fluid}

Thus far, we did not comment on the origin of the jump in the normal stress $[p]$ across the membrane surface \eqref{eq:shape_equation_no_fluid}.
For the case of a passive membrane, $[p]$ captures the jump in the pressure of the surrounding bulk fluid.
In particular, when a lipid membrane fluctuates in a fluid medium, it exerts forces on and experiences forces from the surrounding fluid.
Consider a local shape change in the membrane: the membrane exerts some force on the fluid at that location, the force is transmitted through the fluid, and other regions of the membrane feel a resulting force.
In this section, we first describe how a point force affects the surrounding fluid, and then obtain a dynamical equation which explicitly includes membrane--fluid interactions.

A Newtonian fluid with viscosity $\mu$ acted upon by a point force $\bm{f} \delta (\bmr)$ at location $\bmr := (x, y, z) = \bm{0}$, with negligible inertia, is governed by the Stokes equations
\begin{equation} \label{eq:stokes_equations}
	\nabla \cdot \bmv
	= 0
	\quad \quad
	\text{and}
	\quad \quad
	\mu \nabla^2 \bmv
	- \nabla p
	+ \bm{f} \delta(\bmr)
	= \bm{0}
	~.
\end{equation}
The Green's function solution of the pressure $p$ and velocity $\bmv$ are well-known \cite{sm-Leal} to be given by
\begin{equation} \label{eq:stokes_green_function}
	p(\bmr)
	= \dfrac{\bm{f} \cdot \bmr}{4 \pi r^3}
	\quad \quad
	\text{and}
	\quad \quad
	\bmv(\bmr)
	= \bm{\Lambda} (\bmr) \, \bm{f}
	~,
\end{equation}
where the Oseen tensor $\bm{\Lambda}(\bmr)$ is defined as
\begin{equation} \label{eq:oseen_tensor_def}
	\bm{\Lambda} (\bmr)
	:= \dfrac{1}{8 \pi \mu r} \bigg(
		\bm{I}
		- \dfrac{\bmr \otimes \bmr}{r^2}
	\bigg)
	~.
\end{equation}
Since the membrane deformations are assumed to be small, the forces on the fluid are primarily in the $z$-direction.
Moreover, the resultant pressure and velocity fields at the membrane surface can be approximated by setting $z = 0$ in Eq.~\eqref{eq:stokes_green_function}.
For $\bm{f} = f \bm{e}_z$ and $z = 0$, the fluid pressure $p (x, y, z = 0) = f \delta(x) \delta(y) / 4 \pi$; the fluid velocity is given by
\begin{equation} \label{eq:stokes_velocity_z_zero}
	\bmv (x, y, z = 0)
	= \dfrac{f}{8 \pi \mu \sqrt{x^2 + y^2}} \, \bm{e}_z
	~.
\end{equation}
We also define the $\bm{e}_z \otimes \bm{e}_z$ component of the Oseen tensor at $z = 0$ as
\begin{equation} \label{eq:oseen_scalar_def}
	\Lambda (\bmx)
	:= \dfrac{1}{8 \pi \mu \lvert \bmx \rvert}
	~,
\end{equation}
where
$\bmx = (x, y)$,
such that Eq.~\eqref{eq:stokes_velocity_z_zero} can be equivalently written as
$\bmv (\bmx, 0) = \Lambda(\bmx) \, f \bm{e}_z$.

For a nearly planar lipid membrane in contact with the surrounding fluid, a no-slip boundary condition between the membrane and the bulk fluid can be written as
\begin{equation} \label{eq:no_slip}
    \dfrac{\partial h}{\partial t} (\bmx, t)
    \, = \, v_z (\bmx, z = 0, t)
    \, + \, \eta(\bmx, t)
    ~,
\end{equation}
where $\eta (\bmx, t)$ is a Gaussian random variable capturing perturbations from the surrounding fluid.
Moreover, given a field of point forces per unit area $p(\bmx, z=0, t)$ on the fluid, the $z$-component of the fluid velocity at $z = 0$ is given by
\begin{equation} \label{eq:z_vel_convolution}
    v_z (\bmx, z = 0, t)
    \, = \, \int \mathrm{d} \bmx' \, \Lambda (\bmx - \bmx') \,\, p(\bmx, z = 0, t)
    ~.
\end{equation}
The field $p(\bmx, z=0, t)$ in this case is known to be the force on the fluid by the membrane, which is equal and opposite to the force on the membrane by the fluid---the latter of which is $[p]$.
Thus, according to Eq.\ \eqref{eq:shape_equation_no_fluid},
\begin{equation} \label{eq:p_fluid_passive}
    p(\bmx, z=0, t)
    = - [p]
    = \pint
    = \lambda \nabla^2 h
    - \kappa \nabla^4 h
    ~,
\end{equation}
such that by combining Eqs.\ \eqref{eq:no_slip}--\eqref{eq:p_fluid_passive} we find the dynamical equation governing passive membrane fluctuations is given by \cite{sm-Sapp16, sm-Turlier18} 
\begin{equation} \label{eq:langevin_passive_real}
	\dfrac{\partial h (\bmx, t)}{\partial t}
	= \eta (\bmx, t)
	+ \int \mathrm{d} \bmx' \, \Lambda (\bmx - \bmx') \, \pint (\bmx', t)
	~.
\end{equation}

When characterizing the thermal forces on the membrane from the fluid, as well as when simulating membrane height fluctuations, it is most convenient to work in Fourier space, where the height modes decouple.
To take the Fourier transform of Eq.~\eqref{eq:langevin_passive_real}, we first provide the well-known convolution theorem.
For a general function $f (\bmx, t)$, we have
\begin{equation} \label{eq:convolution_theorem}
	\begin{split}
		\int \mathrm{d} \bmx' \, \Lambda (\bmx - \bmx') \, f (\bmx', t)
		&= \int \mathrm{d} \bmx' \, \dfrac{1}{L} \sum_{\bmk} \hat{\Lambda} (\bmk) \mathrm{e}^{i \bmk \cdot (\bmx - \bmx')} \, f (\bmx', t)
		\\
		&= \sum_{\bmk} \hat{\Lambda} (\bmk) \, \mathrm{e}^{i \bmk \cdot \bmx} \, \dfrac{1}{L} \int \mathrm{d} \bmx' \, f (\bmx', t) \, \mathrm{e}^{-i \bmk \cdot \bmx'}
		\\[4pt]
		&= \sum_{\bmk} \hat{\Lambda} (\bmk) \, \hat{f} (\bmk, t) \, \mathrm{e}^{i \bmk \cdot \bmx}
		~,
	\end{split}
\end{equation}
where in the first line we substituted the Fourier transform of $\Lambda (\bmx - \bmx')$, in the second line we rearranged terms, and in the third line we recognized the form of $\hat{f} (\bmk, t)$.
With the result of Eq.~\eqref{eq:convolution_theorem} and the Fourier transform definitions (\ref{eq:fourier}, \ref{eq:fourier_inverse}), Eq.~\eqref{eq:langevin_passive_real} can be written as
\begin{equation} \label{eq:langevin_passive_fourier_step_1}
	\dfrac{\partial}{\partial t} \bigg(
		\dfrac{1}{L} \sum_{\bmk} \hat{h} (\bmk, t) \, \mathrm{e}^{i \bmk \cdot \bmx}
	\bigg)
	= \dfrac{1}{L} \sum_{\bmk} \hat{\eta} (\bmk, t) \, \mathrm{e}^{i \bmk \cdot \bmx}
	+ \sum_{\bmk} \hat{\Lambda} (\bmk) \, \hat{p}^{\text{int}} (\bmk, t) \, \mathrm{e}^{i \bmk \cdot \bmx}
	~,
\end{equation}
which implies
\begin{equation} \label{eq:langevin_passive_fourier_step_2}
	\dfrac{\partial \hat{h} (\bmk, t)}{\partial t}
	= L \hat{\Lambda} (\bmk) \, \hat{p}^{\text{int}} (\bmk, t)
	+ \hat{\eta} (\bmk, t)
	~.
\end{equation}
The quantities $\hat{\Lambda} (\bmk)$ and $\hat{p}^{\text{int}} (\bmk, t)$ are calculated as
\begin{equation} \label{eq:oseen_p_int_fourier}
	\hat{\Lambda} (\bmk)
	= \dfrac{1}{4 \mu k L}
	\quad \quad
	\text{and}
	\quad \quad
	\hat{p}^{\text{int}} (\bmk, t)
	= - \big(
		\lambda k^2
		+ \kappa k^4
	\big) \, \hat{h} (\bmk, t)
	~,
\end{equation}
such that Eq.~\eqref{eq:langevin_passive_fourier_step_2} can be written as
\begin{equation} \label{eq:langevin_passive_fourier}
	\dfrac{\partial \hat{h} (\bmk, t)}{\partial t}
	= - \omega(k) \, \hat{h} (\bmk, t)
	+ \hat{\eta} (\bmk, t)
	~,
\end{equation}
where the relaxation frequency $\omega(k)$ is given by
\begin{equation} \label{eq:relaxation_frequency}
	\omega(k)
	= \dfrac{1}{4 \mu} \big(
		\lambda k
		+ \kappa k^3
	\big)
	~.
\end{equation}
In Eq.~\eqref{eq:langevin_passive_fourier}, the Fourier transform of the thermal noise, $\hat{\eta} (\bmk, t)$, satisfies the fluctuation--dissipation theorem, such that
\begin{gather}
	\big\langle \hat{\eta}(\bmk, t) \big\rangle
	= 0 ~,
	\label{eq:noise_average}
	\\[9pt]
	\big\langle \textrm{Re}\{\hat{\eta}(\bmk, t)\} \, \textrm{Im}\{\hat{\eta}(\bmk', t')\} \big\rangle
	= 0
	~,
	\label{eq:noise_real_imaginary}
	\\[9pt]
	\big\langle \textrm{Re}\{\hat{\eta}(\bmk, t)\} \, \textrm{Re}\{\hat{\eta}(\bmk', t')\} \big\rangle
	= \kB T L \, \hat{\Lambda}(\bmk) \, \delta(t - t') \, \big(\delta_{\bmk, \bmk'} + \delta_{\bmk, -\bmk'}\big)
	~,
	\label{eq:noise_real_real}
	\\[-3pt]
	\shortintertext{and}
	\big\langle \textrm{Im}\{\hat{\eta}(\bmk, t)\} \, \textrm{Im}\{\hat{\eta}(\bmk', t')\} \big\rangle
	= \kB T L \, \hat{\Lambda}(\bmk) \, \delta(t - t') \, \big(\delta_{\bmk, \bmk'} - \delta_{\bmk, -\bmk'}\big)
	~.
	\label{eq:noise_imaginary_imaginary}
\end{gather}

%
%

\subsection{2.3. Simulation Methodology} \label{sec:sec_passive_simulation_procedure}

In this section, we closely follow the simulation procedure detailed in Ref.~\cite{sm-Sapp16}.
Due to the decoupling of the height modes in Fourier space, each mode is simulated independently.
For a membrane over an $L \times L$ patch with periodic boundary conditions, the allowed wave vectors are
\begin{equation} \label{eq:wave_vectors}
	\bmk
	= (m, n) \, \dfrac{2 \pi}{L}
	~,
	\quad \quad
	m, n \in \mathbb{Z}
	~.
\end{equation}
A space of linearly independent wave numbers, $\mathcal{Q}$, is defined as
\begin{equation} \label{eq:wave_number_space}
	\mathcal{Q}
	= \big\{
		(1 \le m \le M, n = 0)
		\cup
		(0 \le m \le M, 1 \le n \le M)
	\big\}
	~,
\end{equation}
where $M$ defines the largest wave vector considered.
The mode $\bmk = \bm{0}$ is ignored, as it describes only rigid translations of the membrane patch.

To simulate the time evolution of the membrane height modes, Eq.~\eqref{eq:langevin_passive_fourier} is integrated from time $t$ to $t + \Delta t$ to yield
\begin{equation} \label{eq:langevin_passive_fourier_integrated}
	\int_t^{t+\Delta t} \!\!\!\! \mathrm{d}t' \,\, \dfrac{\partial \hat{h}(\bmk, t')}{\partial t'} 
	\, = \, - \omega(k) \int_t^{t + \Delta t} \!\!\!\! \mathrm{d}t' \,\, \hat{h}(\bmk, t') 
	\, + \, \int_t^{t + \Delta t} \!\!\!\! \mathrm{d}t' \,\, \hat{\eta}(\bmk, t')
	~.
\end{equation}
Assuming $\Delta t$ is small, the integrand of the first term on the right-hand side of Eq.~\eqref{eq:langevin_passive_fourier_integrated} is moved outside the integral.
Defining
\begin{equation} \label{eq:langevin_noise_integral}
	\hat{R}(\bmk, t; \Delta t)
	:= \int_t^{t + \Delta t} \!\!\!\! \mathrm{d}t' \,\, \hat{\eta}(\bmk, t')
	~,
\end{equation}
Eq.~\eqref{eq:langevin_passive_fourier_integrated} can be written as
\begin{equation} \label{eq:langevin_passive_fourier_integrated_modified}
	\hat{h}(\bmk, t + \Delta t)
	= \big(
		1 - \omega(k) \Delta t
	\big) \hat{h}(\bmk, t)
	+ \hat{R}(\bmk, t; \Delta t)
	~.
\end{equation}
The complex Gaussian random noise $\hat{R}(\bmk, t; \Delta t)$ has mean zero and variance given by
\begin{equation} \label{eq:fourier_noise_variance}
	\begin{split}
		\big\langle \hat{R}(\bmk, t; \Delta t) \, \hat{R}^*(\bmk, t; \Delta t) \big\rangle
		&= \int_t^{t + \Delta t} \!\!\!\! \mathrm{d}t' \,\, 
		\int_t^{t + \Delta t} \!\!\!\! \mathrm{d}t'' \,\, 
		\big\langle \hat{\eta}(\bmk, t') \hat{\eta}^*(\bmk, t'') \big\rangle
		\\[4pt]
		&= \int_t^{t + \Delta t} \!\!\!\! \mathrm{d}t' \,\, 
		\int_t^{t + \Delta t} \!\!\!\! \mathrm{d}t'' \,\, 
		\bigg(
			\Big\langle \mathrm{Re}\{\hat{\eta}(\bmk, t')\} \, \mathrm{Re}\{\hat{\eta}(\bmk, t'')\} \Big\rangle
			\\[-3pt]
			&\hspace{120pt}+ \Big\langle \mathrm{Im}\{\hat{\eta}(\bmk, t')\} \, \mathrm{Im}\{\hat{\eta}(\bmk, t'')\} \Big\rangle
		\bigg)
		\\[5pt]
		&= 2 \kBT L \hat{\Lambda}(\bm{k}) \Delta t
		~,
	\end{split}
\end{equation}
where in the first equality Eq.~\eqref{eq:langevin_noise_integral} was substituted, in the second equality $\hat{\eta}$ was split into real and imaginary parts and Eq.~\eqref{eq:noise_real_imaginary} was used to eliminate cross terms, and in the third equality Eqs.~\eqref{eq:noise_real_real} and \eqref{eq:noise_imaginary_imaginary} were substituted.
Defining $r_1$ and $r_2$ to be independent, normally distributed random numbers, the height modes are evolved numerically according to
\begin{equation} \label{eq:langevin_passive_fourier_discretized}
	\hat{h}(\bmk, t + \Delta t)
	= \big(
		1 - \omega(k) \Delta t
	\big) \hat{h}(\bmk, t)
	+ \big(
		r_1
		+ i \, r_2
	\big) \sqrt{\kBT L \hat{\Lambda}(\bm{k}) \Delta t} 
	~.
\end{equation}
Note that in Eq.~\eqref{eq:langevin_passive_fourier_discretized}, $r_1$ and $r_2$ are used to distribute the random noise in both the real and imaginary directions, each with a variance of one-half the result of Eq.~\eqref{eq:fourier_noise_variance}.
In practice, the real and imaginary components of the height modes are simulated independently.
Our code to calculate the fluctuation spectrum by evolving height modes according to Eq.\ \eqref{eq:langevin_passive_fourier_discretized} is provided at \texttt{\href{https://github.com/mandadapu-group/active-contact}{https://github.com/mandadapu-group/active-contact}}.

%
%

\subsection{2.4. Theoretical, Numerical, and Experimental Results} \label{sec:sec_passive_results}

\begin{figure}[t!]
	\centering
	\subfigure[\ passive, no average]{\includegraphics[width=0.30\linewidth]{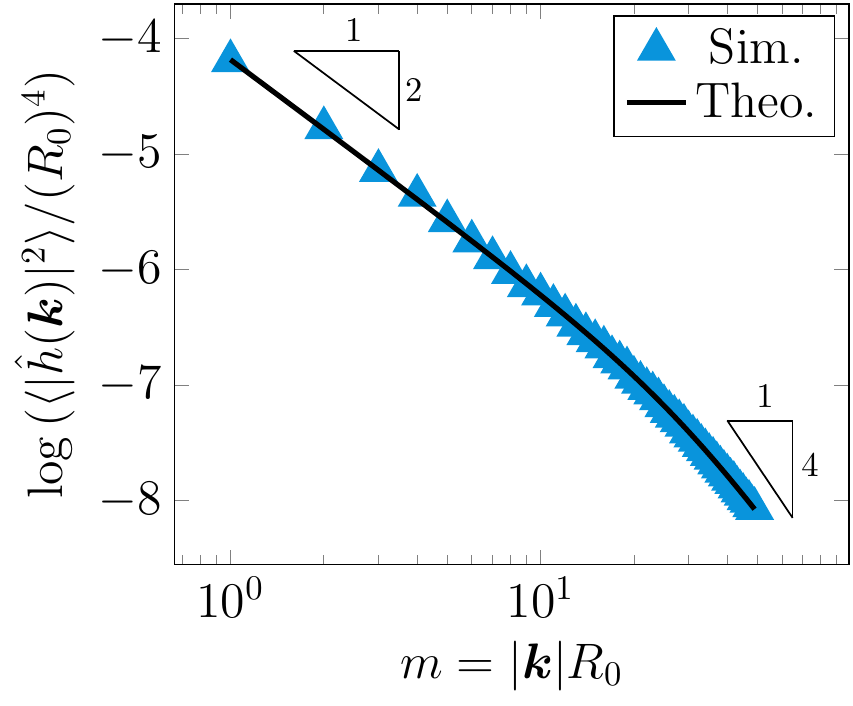}
	\label{fig:fig_results_passive_no_avg}}
	\subfigure[\ passive, average]{\includegraphics[width=0.30\linewidth]{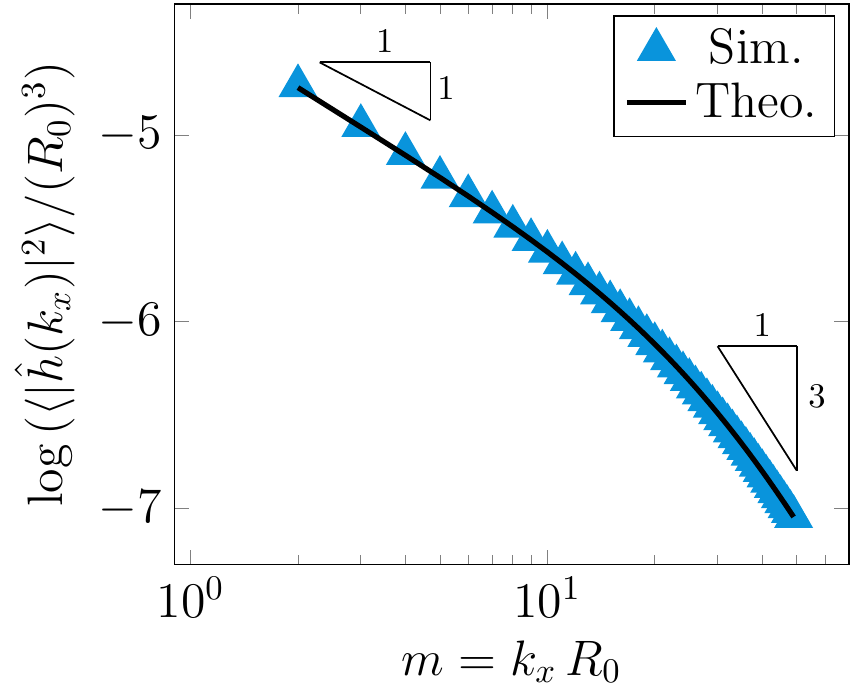}
	\label{fig:fig_results_passive_avg}}
	\subfigure[\ passive, average, with data]{\includegraphics[width=0.30\linewidth]{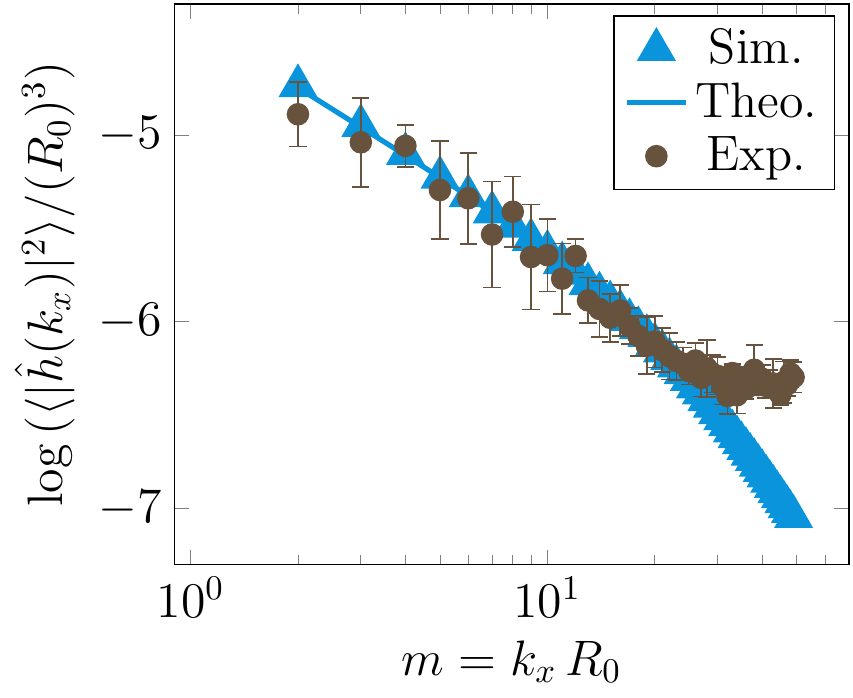}
	\label{fig:fig_results_passive_avg_data}}
	\caption{
		Passive thermal fluctuations of a lipid membrane in thermal equilibrium with the surrounding fluid.
		(a) Comparison of dynamical simulations, as described in Sec.\ 2.3 (blue triangles) and the known equilibrium result of Eq.~\ref{eq:helfrich_fluctuation_spectrum} (black curve).
		The quantitative agreement indicates the code is working as expected.
		(b) The result of averaging the simulation result and theoretical prediction over $k_y$ modes, according to Eq.~\eqref{eq:fluctuation_averaging_numerical}.
		(c) Experimental passive data overlaid on the averaged passive result.
		The systematic discrepancy at large $k_x$ occurs due to the pixel resolution of the camera.
		In all simulations and presented theoretical results, parameters are
		$\lambda = 4 \cdot 10^{-3}$ pN/nm,
		$\kappa = 14.3 \, \kBT$ at $T = 30^\circ$C,
		$R_0$ $ = 4 \, \mu$m,
		$\mu = 0.7972$ mPa$\cdot$s,
		and height fluctuations were simulated over 0.7 s.
	}
	\label{fig:fig_results_passive}
\end{figure}

We now present the results of passive numerical simulations to
(i) show the numerical scheme reproduces equilibrium fluctuations, and
(ii) demonstrate how simulations are compared to experiments.
For each wave vector $\bmk$, the simulations generate $\text{Re} \{ \hat{h} (\bmk, t) \}$ and $\text{Im} \{ \hat{h} (\bmk, t) \}$ over time, with which
$
	\big \langle
		\big \lvert \hat{h} (\bmk, t) \big \rvert^2
	\big \rangle_{\text{pas}}
$
is calculated.
As shown in Fig.~\ref{fig:fig_results_passive_no_avg}, the simulations (blue triangles) exactly match the known theoretical result (Eq.~\eqref{eq:helfrich_fluctuation_spectrum}, black line).

As vesicles are imaged experimentally at a single cross-section, all Fourier modes orthogonal to this cross-section are implicitly summed over.
To compare simulation results with experiments, the height fluctuations of the nearly planar membrane are averaged over $k_y$ modes according to Eq.~\eqref{eq:fluctuation_averaging}.
In practice, the averaging is done numerically, according to
\begin{equation} \label{eq:fluctuation_averaging_numerical}
	\big \langle
		\big \lvert \hat{h} (k_x, t) \big \rvert^2
	\big \rangle
	= \dfrac{2}{L} \sum_{n = 0}^M
		\big \langle
			\big \lvert \hat{h} \big(
				\bmk
				= (k_x, 2 \pi n / L),
				t
			\big) \big \rvert^2
		\big \rangle
	~.
\end{equation}
Moreover, the length $L$ in simulations is set to
$L = 2 \pi R_0$,
where $R_0$ is the radius of the undeformed membrane vesicle, to be consistent with the Fourier transform of experimental data (see Eq.~\eqref{eq:expFT}).
In averaging our simulation results according to Eq.\ \eqref{eq:fluctuation_averaging_numerical}, we obtain the results shown as blue triangles in Fig.~\ref{fig:fig_results_passive_avg}, which agree with the theoretical calculation of Eq.\ \eqref{eq:helfrich_fluctuation_spectrum_averaged} (black line).
In Fig.\ \ref{fig:fig_results_passive_avg_data}, the data contained in Fig.\ \ref{fig:fig_results_passive_avg} are overlaid with experimental data.
Figure~\ref{fig:fig_results_passive_avg_data} contains the same information as the passive portion of Fig.~2 in the main text, following the same color scheme.

%
%

\section{3. Theory and Simulation of Active Membranes} \label{sec:sec_active_membranes}

In this section, the non-equilibrium theory and simulations of Sec.\ 2 are extended to model lipid membrane vesicles acted upon by active bacterial contact forces.
When bacteria push on the membrane surface, a new force enters the membrane shape equation, which in turn is transmitted throughout the fluid to exert forces at other locations on the membrane surface.
Importantly, we spread the bacterial contact force over the width of a bacterium, and recognize the characteristic duration of bacterial--membrane contact is much larger than the timescale of membrane fluctuations, $1 / \omega(q)$.
As a result, the membrane fluctuation spectrum can be written as the sum of two terms: an equilibrium term identical to that of a passive membrane, and an active term involving details of the bacterial contact force.

%
%

\subsection{3.1. Non-Equilibrium Contact Theory} \label{sec:sec_active_nonequilibrium_theory}

With a model for the dynamical height fluctuations of a passive membrane vesicle, we now seek to describe the active membrane fluctuations resulting from self-propelled bacteria contained within a membrane vesicle.
The active particles exert a force on the membrane, which we approximate by the active force per area
\begin{equation} \label{eq:active_force_real}
	\pact
	= \sum_{j = 1}^{\Nc}
		\bar{p} \, \phi(t; t_j) \, \exp \bigg\{
			- \dfrac{(\bmx - \bmx_j)^2}{2 a^2}
		\bigg\}
	~.
\end{equation}
In Eq.~\eqref{eq:active_force_real}, $\Nc$ is the number of collision events, with the $j^{\text{th}}$ active particle--membrane collision occurring at time $t_j$ and position $\bmx_j$.
The only dimensional quantity on the right-hand side of Eq.~\eqref{eq:active_force_real} is $\bar{p}$, which captures the maximum pressure exerted by the particle on the membrane.
As a simple approximation, we set
$\bar{p} = 2 \lambda / a$,
where $a$ is the half-width of a bacterium and $\bar{p}$ would be the pressure exerted by a membrane on a sphere of radius $a$.
The Gaussian contribution in Eq.\ \eqref{eq:active_force_real} describes the spreading of the particle--membrane contact point force over an area.
Lastly, $\phi(t; t_j)$ approximates the temporal nature of the particle--membrane collision.
As shown in Fig. 3b in the main text, $\phi (t; t_j)$ is an isosceles trapezoid centered at time $t_j$ with top length $\tauR$ and bottom length $\tauR + 2 \tauP$;
$\tauP \approx 0.05$ sec
is an estimate of how long it takes for a bacterium to come to a complete stop due to elastic membrane forces, once it makes initial contact with the membrane.

With a characterization of the active forces on the membrane, we follow an identical procedure to that of the passive case.
The jump in the normal traction acting on the membrane is now given by
$[p] = -\ptot + \pact$, such that
the magnitude of the total force per area $\ptot$ acting on the membrane by the surrounding fluid can be written as
\begin{equation} \label{eq:total_force_real}
	\ptot (\bmx, t)
	= \pint (\bmx, t)
	+ \pact (\bmx, t)
	~.
\end{equation}
Recognizing $p(\bmx, z=0, t) = \ptot$ (c.f.\ Eqs.\ \eqref{eq:z_vel_convolution} and \eqref{eq:p_fluid_passive}), we find the active analog of Eq.~\eqref{eq:langevin_passive_real} is given by
\begin{equation} \label{eq:langevin_active_real}
	\dfrac{\partial h (\bmx, t)}{\partial t}
	= \eta (\bmx, t)
	+ \int \mathrm{d} \bmx' \, \Lambda (\bmx - \bmx') \, \Big[
	    \pint (\bmx', t)
	    + \pact (\bmx', t)
	\Big]
	~.
\end{equation}
Again taking the Fourier transform of Eq.~\eqref{eq:langevin_active_real} and using the convolution theorem~\eqref{eq:convolution_theorem}, we obtain
\begin{equation} \label{eq:langevin_active_fourier_step_1}
	\dfrac{\partial \hat{h} (\bmk, t)}{\partial t}
	= - \omega(k) \, \hat{h} (\bmk, t)
	+ \hat{\eta} (\bmk, t)
	+ L \hat{\Lambda} (\bmk) \, \hatpact (\bmk, t)
	~,
\end{equation}
where the Fourier transform of the active pressure is calculated to be
\begin{equation} \label{eq:active_pressure_fourier}
	\hatpact (\bmk, t)
	= \, \sum_{j = 1}^{\Nc}
	\dfrac{a^2 \, \bar{p}}{R_0} \,
	\phi(t; t_j) \,
	\exp \Big\{
		-i \bmx_j \cdot \bmk
		- \dfrac{a^2 \, k^2}{2}
	\Big\}
	~.
\end{equation}
In Eq.~\eqref{eq:active_pressure_fourier}, we substituted $L = 2 \pi R_0$ to simplify the expression.
By substituting Eqs.~\eqref{eq:oseen_p_int_fourier}$_1$ and \eqref{eq:active_pressure_fourier} into Eq.~\eqref{eq:langevin_active_fourier_step_1}, we obtain
\begin{equation} \label{eq:sm_langevin_active_fourier}
	\dfrac{\partial \hat{h} (\bmk, t)}{\partial t}
	\, = \, - \omega(k) \, \hat{h} (\bmk, t)
	\, + \, \hat{\eta} (\bmk, t)
	\, + \, \sum_{j = 1}^{\Nc}
	\dfrac{a^2 \, \bar{p}}{4 \mu k R_0} \,
	\phi(t; t_j) \,
	\exp \Big\{
		-i \bmx_j \cdot \bmk
		- \dfrac{a^2 \, k^2}{2}
	\Big\}
	~.
\end{equation}
Equation~\eqref{eq:sm_langevin_active_fourier} is presented as Eqs.~(3) and (4) in the main text.
As discussed in the main text, an approximate solution of the height fluctuation spectrum given by Eq.\ \eqref{eq:sm_langevin_active_fourier} is found to be
\begin{equation} \label{eq:fluctuation_theory_sm}
	\langle \lvert \hat{h} (\bm{k}) \rvert^2 \rangle
	\, = \, \dfrac{\kBT}{\kappa k^4 + \lambda k^2}
	\, + \, \dfrac{N_{\mathrm{p}} \, \tauR}{\tauT + \tauR}
	\bigg(
		\dfrac{a^2 \, \bar{p} / R_0}{\kappa k^4 + \lambda k^2}
	\bigg)^{\!\! 2} \, \mathrm{e}^{ - a^2 k^2 }
	~,
\end{equation}
where $N_{\mathrm{p}}$ is the number of enclosed bacteria, $\tauR$ is the bacteria reorientation time, and $\tauT$ is the time it takes the bacteria to travel from one side of the vesicle to the other.

%
%

\subsection{3.2. Simulation Methodology} \label{sec:sec_active_simulation_procedure}

Just as the active non-equilibrium theory is an extension of its passive analog, we extend the passive simulation methodology to simulate lipid membrane vesicles being acted upon by active contact forces.
By integrating Eq.~\eqref{eq:sm_langevin_active_fourier} from time $t$ to $t + \Delta t$ and recognizing only the active pressure term is new, we find the height modes are evolved according to
\begin{equation} \label{eq:langevin_active_fourier_discretized}
	\begin{split}
		\hat{h}(\bmk, t + \Delta t)
		= \big(
			1 - \omega(k) \Delta t
		\big) \hat{h}(\bmk, t)
		&+ \big(
			r_1
			+ i \, r_2
		\big) \sqrt{\kBT L \hat{\Lambda}(\bm{k}) \Delta t} 
		\\[6pt]
		&+ \sum_{j = 1}^{\Nc}
		\dfrac{a^2 \, \bar{p} \, \Delta t}{4 \mu k R_0} \,
		\phi(t; t_j) \,
		\exp \Big\{
			-i \bmx_j \cdot \bmk
			- \dfrac{a^2 \, k^2}{2}
		\Big\}
		~.
	\end{split}
\end{equation}
As before, the real and imaginary components of each membrane mode is simulated independently.
In code, the number of collisions
$\Nc = \Np \cdot \tsim / (\tauR + \tauT)$,
where  $\Np = 7$ is the number of particles, $\tsim = 7$ sec is the total simulation time, $\tauR = 0.5$ sec is the bacterial reorientation time, and $\tauT = 0.5$ sec is the traversal time---the latter of which is the time it takes for the bacteria to go from one end of the vesicle to another, given the bacterial swim speed $U_0 \approx 15 \, \mu$m/s.
Moreover, the collision times $t_j$ and position $\bmx_j$ are chosen randomly from a uniform distribution of times in the range $[0, \tsim]$ and positions in the range $[0, L] \times [0, L]$, respectively.
Again, our code is provided at \texttt{\href{https://github.com/mandadapu-group/active-contact}{https://github.com/mandadapu-group/active-contact}}.

%
%

\subsection{3.3. Results} \label{sec:sec_active_results}

\begin{figure}[t!]
	\centering
	\subfigure[\ active, average]{\includegraphics[width=0.34\linewidth]{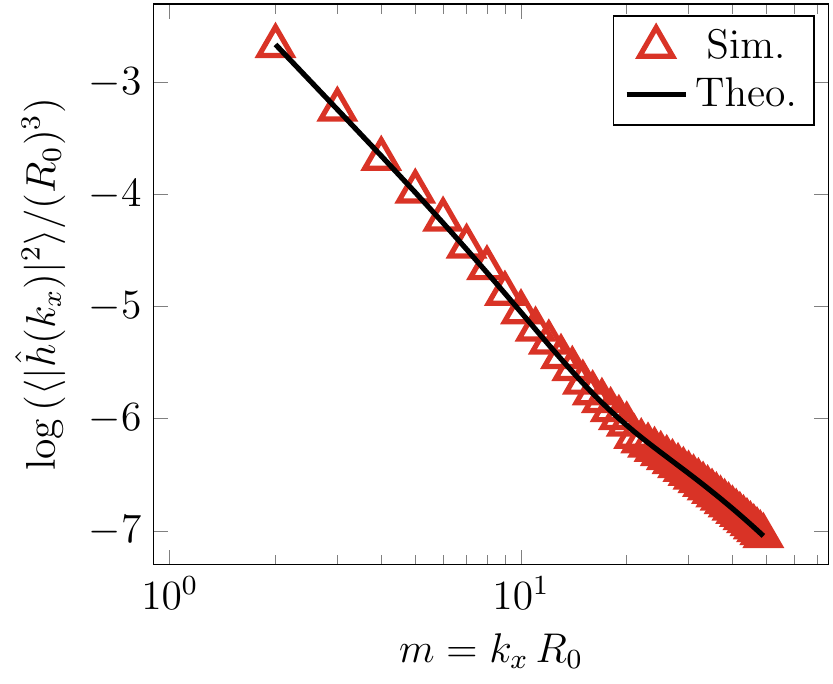}
	\label{fig:fig_results_active_avg}}
	\subfigure[\ active, average, with data]{\includegraphics[width=0.34\linewidth]{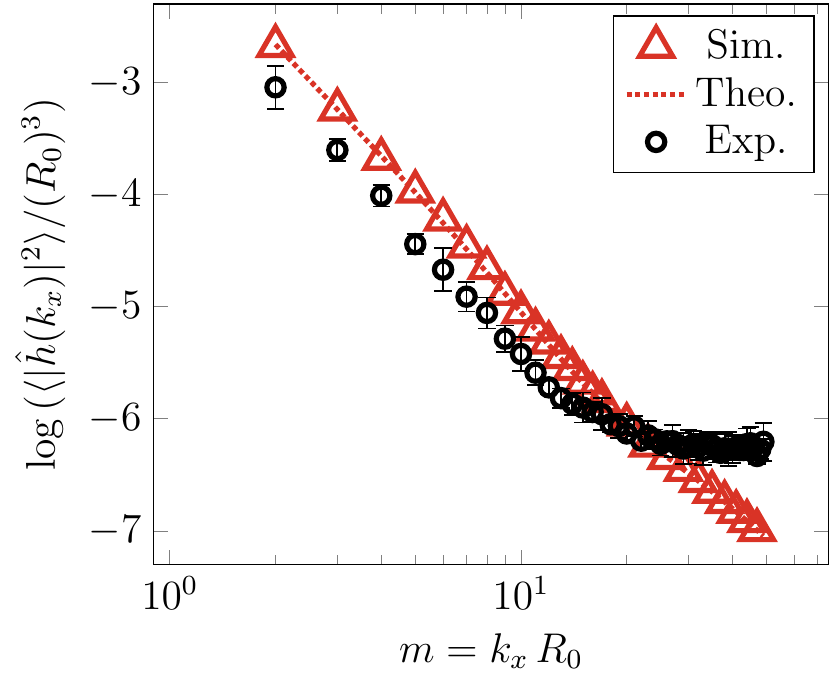}
	\label{fig:fig_results_active_avg_data}}
	\caption{
		Active lipid membrane fluctuations.
		(a) Simulation results (red triangles) show excellent agreement with the theoretical prediction (black curve, expression in main text).
		(b) Experimental data overlaid on the same plot.
		Again, the system leveling off of the experimental fluctuations at large $k_x$ occurs due to camera resolution and the intrinsic noise present at large wave vectors.
		All simulation parameters are identical to those detailed in Fig.~\ref{fig:fig_results_passive}, and additional details can be found in our code---provided at
		\texttt{\href{https://github.com/mandadapu-group/active-contact}{https://github.com/mandadapu-group/active-contact}}.
	}
	\label{fig:fig_results_active}
\end{figure}

\begin{figure}[t!]
	\centering
	\subfigure[\ Active fluctuations: $R_0 \approx 8$ $\mu$m, $\Np \approx 10$] {\includegraphics[width=0.36\linewidth]{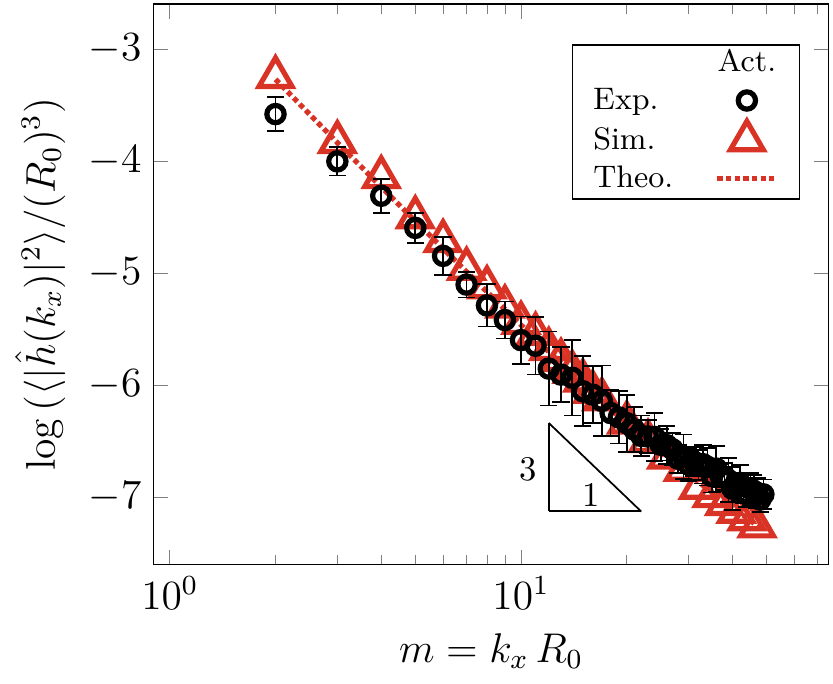}
	\label{fig:Bsub_addl_8um}}		
	\hspace{20pt}
	\subfigure[\ Active fluctuations: $R_0 \approx 15$ $\mu$m, $\Np \approx 20$] {\includegraphics[width=0.36\linewidth]{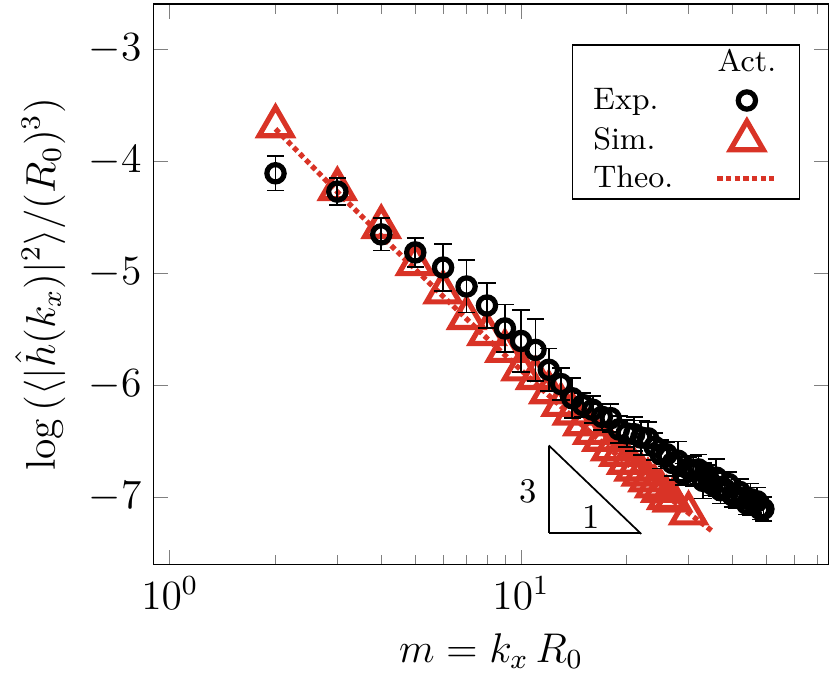}
	\label{fig:Bsub_addl_15um}}
	\subfigure[\ Snapshot of a GUV (radius $\approx 15$ $\mu$m) containing $\approx 20$ motile bacteria, which are visible as black objects in the brightfield image. ]{\includegraphics[width=0.27\linewidth]{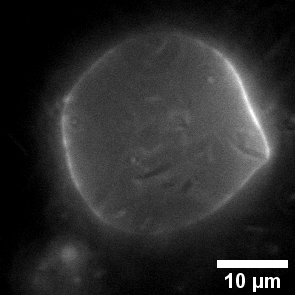}
	\label{fig:Bsub17}}
	\subfigure[\ Fluorescence image of the same GUV at a different time, which is used for height fluctuation analysis. ]{\includegraphics[width=0.27\linewidth]{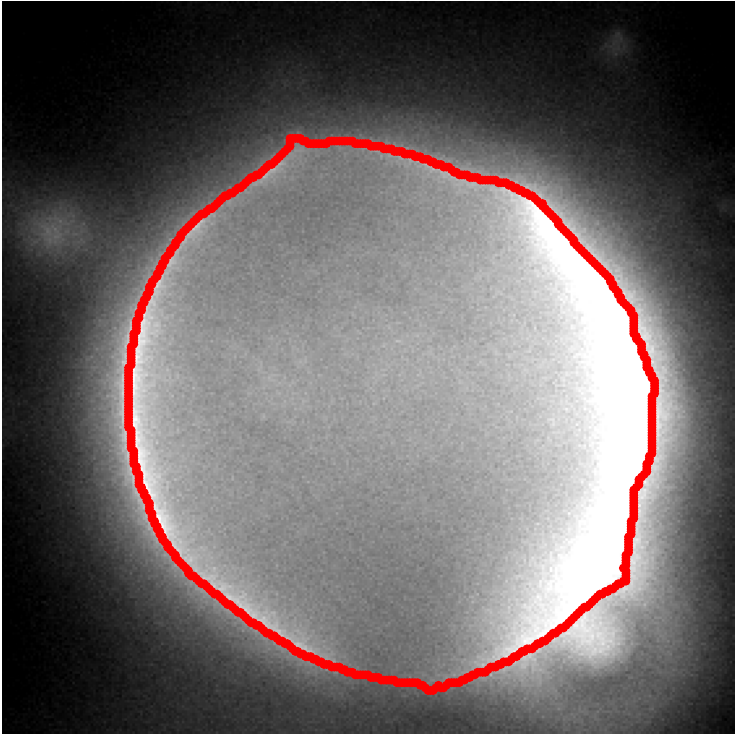}
	\label{fig:Bsub16_contour}}
	\subfigure[\ Deformation of the membrane as a function of angle around the GUV.]{\includegraphics[width=0.33\linewidth]{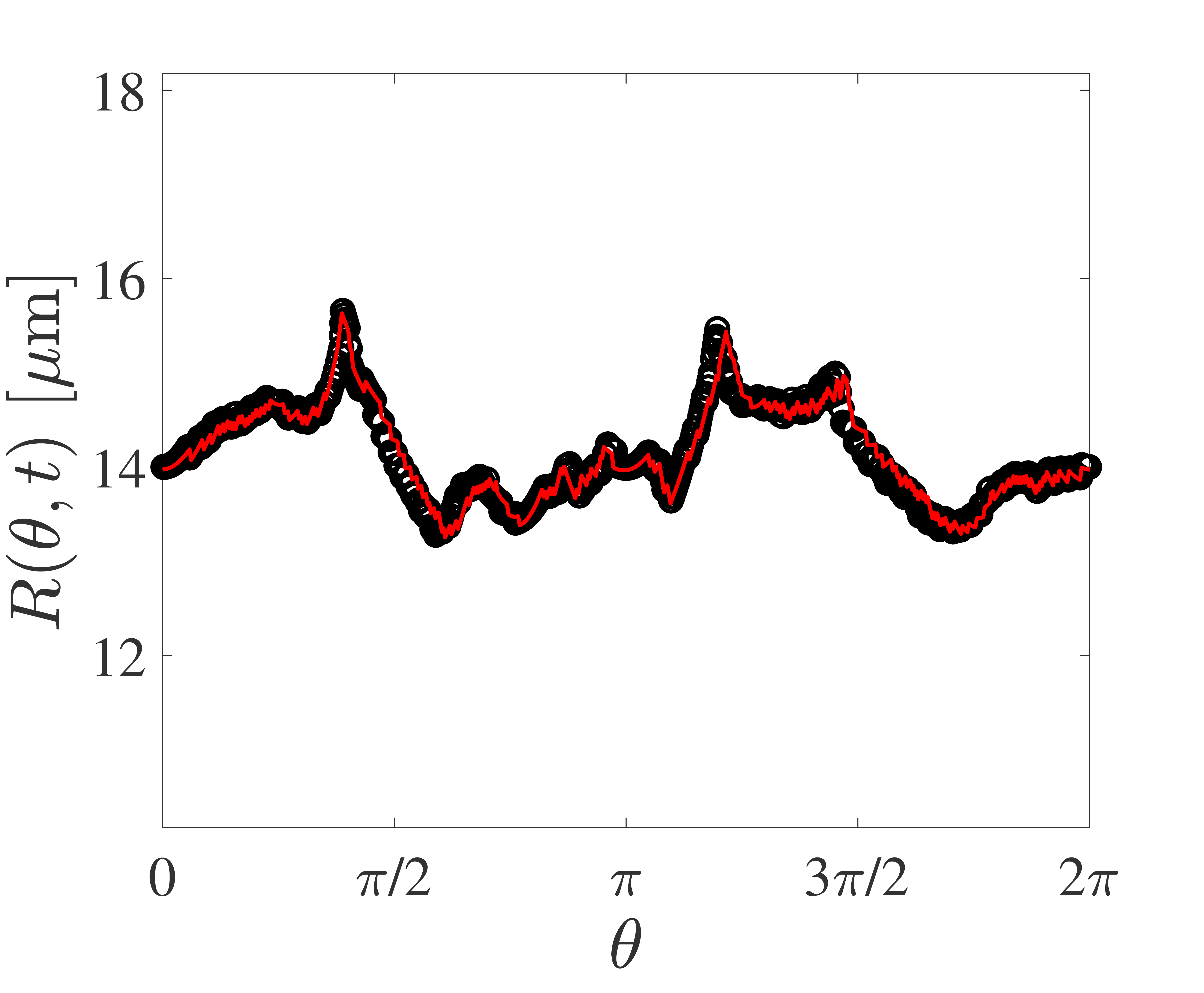}
	\label{fig:Bsub16_contour_image}}
	\caption{Experimental data for shape fluctuations of bacteria-containing vesicles, in two cases: (a) GUV with radius $R_0 \approx 8$ $\mu$m and $\Np \approx 10$ bacteria, and (b) GUV with radius $R_0 \approx 15$ $\mu$m containing $\Np \approx$ 20 bacteria.
	(c)--(d) Instantaneous snapshots of the vesicle corresponding to case (b), via brightfield and fluorescence imaging.  In (d), the red curve denotes the membrane contour, calculated using an edge-detection algorithm. (e) Membrane deformation as a function of arclength around the membrane perimeter.  Black symbols are data and the red curve is the Fourier spectra to obtain the height fluctuations.  }
	\label{fig:additional_exp_1}
\end{figure}

As shown in Fig.\ \ref{fig:fig_results_active_avg}, there is excellent agreement between our simulation results and the theoretical prediction of Eq. \eqref{eq:fluctuation_theory_sm}.
Note that Fig.\ \ref{fig:fig_results_active} contains the same data as was presented in the main text, for which $\Np \approx 7$ and $R_0 \approx 4$ $\mu$m.
To test the robustness of our theoretical model, we now also provide an analysis of two additional active vesicles, as shown in Fig.~\ref{fig:additional_exp_1}: one with $\Np \approx 10$ and $R_0 \approx 8$ $\mu$m, and another with $\Np \approx 20$ and $R_0 \approx 15$ $\mu$m.
In these experiments, the passive data was not available, and so the surface tension and bending modulus for these vesicles could not be obtained.
In our analysis, we used the values of $\lambda$ and $\kappa$ from the 7-particle case.
However, as we show in the following section, our theoretical results are insensitive to the values of $\lambda$ and $\kappa$, and so we still obtain reasonable predictions given this limitation.

As seen in Fig.~\ref{fig:Bsub_addl_8um}, the 10-particle vesicle again shows excellent agreement between experiments, simulation, and theory, thus demonstrating the validity of our numerical and analytical developments.  
The results from the 20-particle vesicle, on the other hand, suggest where our theory begins to break down.
As shown in Fig.~\ref{fig:Bsub_addl_15um}, although there is generally good agreement with the experimental data, there is a slight difference in the shape of the latter: active fluctuations at lower modes are slightly suppressed, while active fluctuations at intermediate modes are slightly enhanced.  We believe this qualitative change is due to there being more bacteria enclosed within the vesicle. 
As can be seen from Figs.~\ref{fig:Bsub17}--\ref{fig:Bsub16_contour_image}, there are now often times where multiple bacteria contact a local portion of the membrane in quick succession.
Due to their persistent motion, active particles have a tendency to accumulate at surfaces \cite{sm-yan2015force, sm-nikola2016active}, and it seems that in the 20-particle vesicle such effects are no longer negligible.
Importantly, when multiple bacteria contact nearby regions of the membrane in rapid succession, large wavelength modes are effectively converted into shorter wavelength ones, as can be seen by comparing Fig.~\ref{fig:fig_edge_detection} with Fig.~\ref{fig:Bsub16_contour_image}.
In the former, the membrane receives isolated, single perturbations that relax fully before the membrane receives the next active perturbation, while in the latter, there is a superposition of many active perturbations which occur simultaneously---effectively decreasing the magnitude of low modes and increases the magnitude of intermediate ones.
We thus find that while our theory captures the shape fluctuations of active membranes a cross a range of vesicle sizes and active particle numbers, it is most accurate when particle numbers are low and bacteria--bacteria correlations do not significantly affect bacteria--membrane interactions.

%
%

\subsection{3.4 Parameter sensitivity analysis of theoretical model}

\begin{figure}[t!]
	\centering
	\subfigure[\ bending modulus $\kappa$]
	{\includegraphics[width=0.32\linewidth]{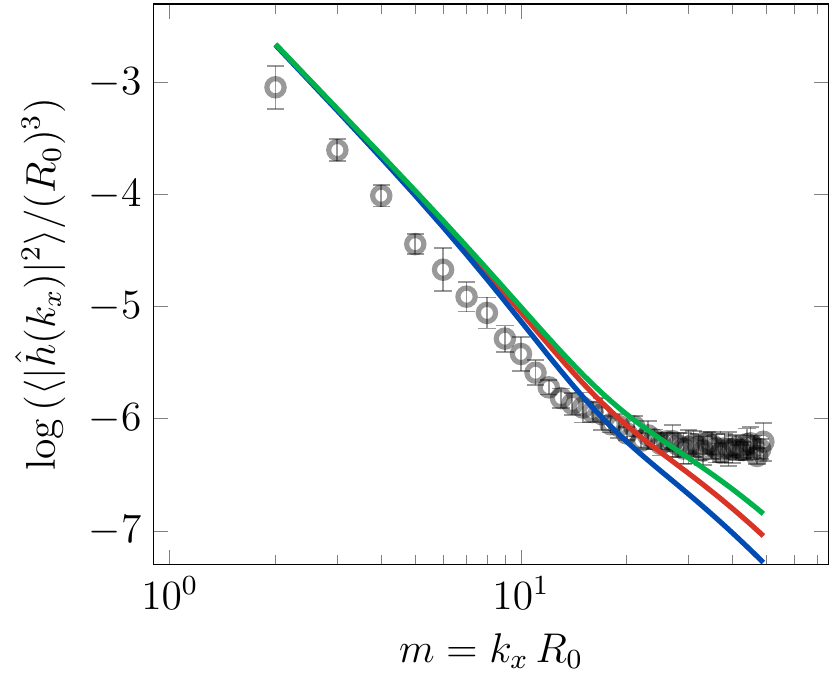}
	\label{fig:kappa_sensitivity}}
	\subfigure[\ surface tension $\lambda$]
	{\includegraphics[width=0.32\linewidth]{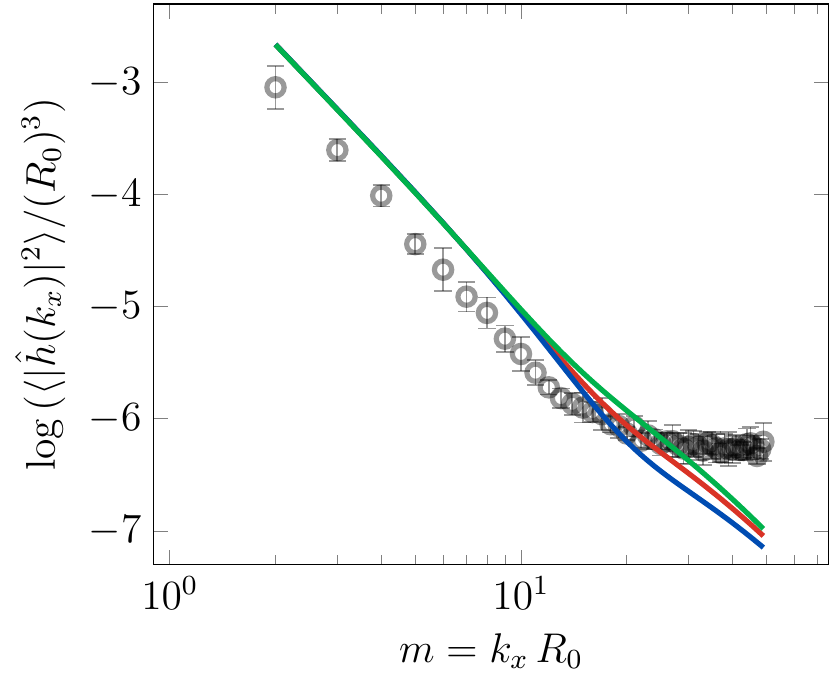}
	\label{fig:lambda_sensitivity}}
	\subfigure[\ reorientation time $\tauR$]
	{\includegraphics[width=0.32\linewidth]{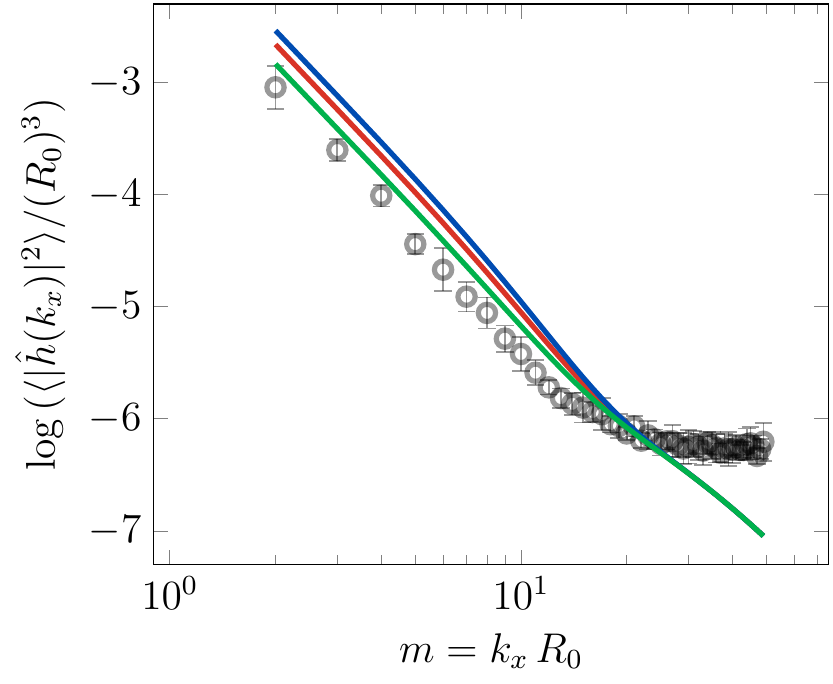}
	\label{fig:tauR_sensitivity}}
	\subfigure[\ swim speed $U_0$]
	{\includegraphics[width=0.32\linewidth]{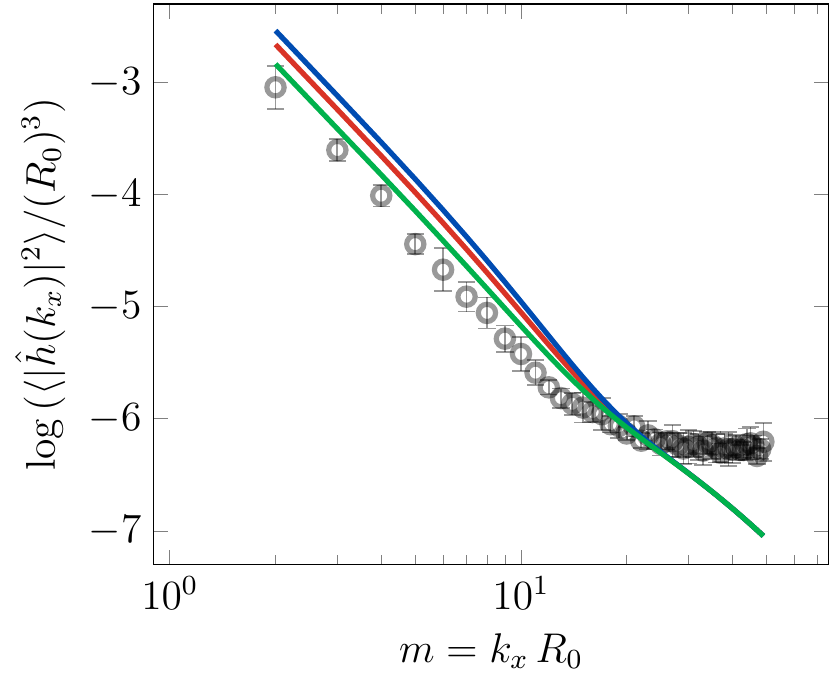}
	\label{fig:vel_sensitivity}}		
	\subfigure[\ half-width $a$]
	{\includegraphics[width=0.32\linewidth]{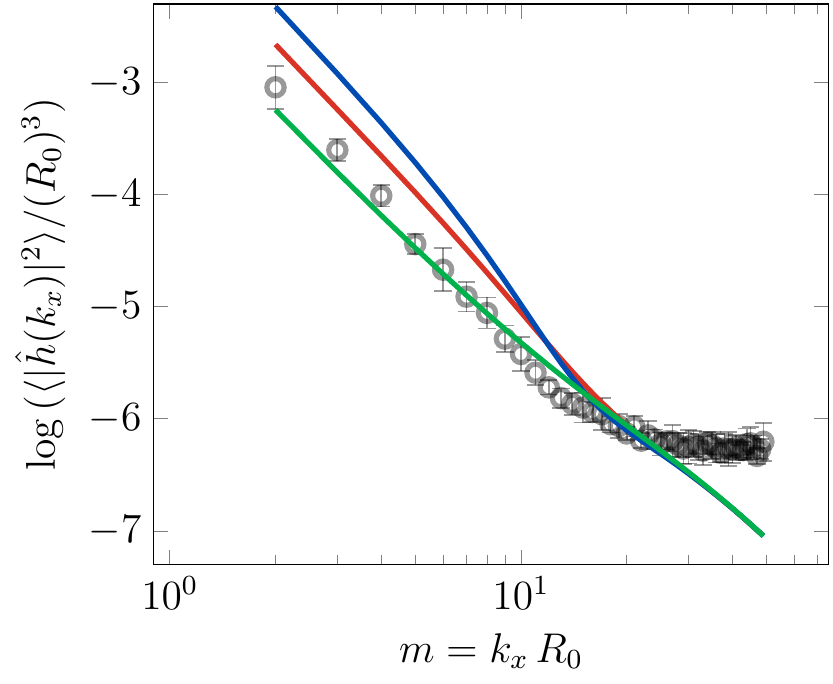}
	\label{fig:arad_sensitivity}}
	\caption{
    	Sensitivity analysis of the theoretical result, Eq.\ \eqref{eq:fluctuation_theory_sm}, to changes in parameter values.
    	In all cases, the red line is the theoretical result presented in Fig.\ 2 of the main text, and the open circles are the active fluctuation data.
    	For plots (a)--(d), the green line represents a decrease in the chosen parameter by a factor of two, while the blue line represents an increase in the chosen parameter by a factor of two.
    	For plot (e), as the analytical expression is sensitive to the bacterial half-width through the exponential term, the green line represents a decrease in the bacterial half-width by 50\%, while the blue line represents an increase in the bacterial half-width by 50\%.
    }
	\label{fig:sensitivity}
\end{figure}

In considering the experimental system, there are seven fundamental parameters: the bending modulus $\kappa$, surface tension $\lambda$, vesicle radius $R_0$, number of bacteria $\Np$, bacterial reorientation time $\tauR$, bacterial half-width $a$, and bacterial velocity $U_0$.
From these, we approximate the magnitude of the contact pressure as $\bar{p} \approx 2 \lambda / a$ and the bacterial traversal time $\tauT \approx 2 R_0 / U_0$.
We have already experimentally demonstrated how changes to $\Np$ and $R_0$ alter the active fluctuation spectrum, however the remaining parameters are not easily modified experimentally.
Thus, we understand how our theoretical results would change due to variations in the remaining fundamental parameters via a sensitivity analysis.
As shown in Fig.\ \ref{fig:sensitivity}, our analytical prediction is relatively sensitive to changes in the bacterial half-width $a$, but otherwise fairly insensitive to changes in the remaining parameters.

%
%

\section{4. Supplemental Videos}

Below, we describe the Supplemental Videos associated with this manuscript.  In all movies, the time stamp corresponds to minutes:seconds.

\begin{enumerate}
	\item[] \textbf{S1.} Fluorescence movie of a giant unilamellar vesicle (GUV) containing several non-motile \textit{B.\ subtilis}.
	
	\item[] \textbf{S2.} Brightfield movie of a GUV containing several motile \textit{B.\ subtilis}. The vesicle edges can be seen as a thin black line. 
	
	\item[] \textbf{S3.} Fluorescence movie of the same GUV as in Vid.\ S2, containing several motile \textit{B.\ subtilis}. Bacteria are non-fluorescent and are not visible in this movie. 
	
	\item[] \textbf{S4.} Merged fluorescence and brightfield movie of the same GUV containing several motile \textit{B.\ subtilis}.
	
	\item[] \textbf{S5.} Merged fluorescence and brightfield movie of a floppy GUV containing motile \textit{B.\ subtilis}. Membrane deformations are larger for this GUV.
	
	\item[] \textbf{S6.} Merged fluorescence and brightfield movie of a GUV containing Janus particles in the absence of hydrogen peroxide.  The scale bar is 10 $\mu$m.
	
	\item[] \textbf{S7.}  Merged fluorescence and brightfield movie of a GUV containing Janus particles in the presence of 0.5\% hydrogen peroxide.  Self propulsion of the Janus particles can be observed and their collisions with the membrane. 
	
	\item[] \textbf{S8.} Merged fluorescence and brightfield movie of a GUV containing many motile \textit{B.\ subtilis}. Deformations are very large and thin membrane tubes can be seen. Each membrane tube contain a few bacteria that collided into the membrane.
\end{enumerate}

%% file: sm.bbl
%

%% file: main.bbl
\begin{thebibliography}{60}%
\makeatletter
\providecommand \@ifxundefined [1]{%
 \@ifx{#1\undefined}
}%
\providecommand \@ifnum [1]{%
 \ifnum #1\expandafter \@firstoftwo
 \else \expandafter \@secondoftwo
 \fi
}%
\providecommand \@ifx [1]{%
 \ifx #1\expandafter \@firstoftwo
 \else \expandafter \@secondoftwo
 \fi
}%
\providecommand \natexlab [1]{#1}%
\providecommand \enquote  [1]{``#1''}%
\providecommand \bibnamefont  [1]{#1}%
\providecommand \bibfnamefont [1]{#1}%
\providecommand \citenamefont [1]{#1}%
\providecommand \href@noop [0]{\@secondoftwo}%
\providecommand \href [0]{\begingroup \@sanitize@url \@href}%
\providecommand \@href[1]{\@@startlink{#1}\@@href}%
\providecommand \@@href[1]{\endgroup#1\@@endlink}%
\providecommand \@sanitize@url [0]{\catcode `\\12\catcode `\$12\catcode
  `\&12\catcode `\#12\catcode `\^12\catcode `\_12\catcode `\%12\relax}%
\providecommand \@@startlink[1]{}%
\providecommand \@@endlink[0]{}%
\providecommand \url  [0]{\begingroup\@sanitize@url \@url }%
\providecommand \@url [1]{\endgroup\@href {#1}{\urlprefix }}%
\providecommand \urlprefix  [0]{URL }%
\providecommand \Eprint [0]{\href }%
\providecommand \doibase [0]{http://dx.doi.org/}%
\providecommand \selectlanguage [0]{\@gobble}%
\providecommand \bibinfo  [0]{\@secondoftwo}%
\providecommand \bibfield  [0]{\@secondoftwo}%
\providecommand \translation [1]{[#1]}%
\providecommand \BibitemOpen [0]{}%
\providecommand \bibitemStop [0]{}%
\providecommand \bibitemNoStop [0]{.\EOS\space}%
\providecommand \EOS [0]{\spacefactor3000\relax}%
\providecommand \BibitemShut  [1]{\csname bibitem#1\endcsname}%
\let\auto@bib@innerbib\@empty
\bibitem [{\citenamefont {Lewis}\ \emph {et~al.}(1996)\citenamefont {Lewis},
  \citenamefont {Rousso}, \citenamefont {Khachatryan}, \citenamefont {Brodsky},
  \citenamefont {Lieberman},\ and\ \citenamefont {Sheves}}]{lewis1996directly}%
  \BibitemOpen
  \bibfield  {author} {\bibinfo {author} {\bibfnamefont {A.}~\bibnamefont
  {Lewis}}, \bibinfo {author} {\bibfnamefont {I.}~\bibnamefont {Rousso}},
  \bibinfo {author} {\bibfnamefont {E.}~\bibnamefont {Khachatryan}}, \bibinfo
  {author} {\bibfnamefont {I.}~\bibnamefont {Brodsky}}, \bibinfo {author}
  {\bibfnamefont {K.}~\bibnamefont {Lieberman}}, \ and\ \bibinfo {author}
  {\bibfnamefont {M.}~\bibnamefont {Sheves}},\ }\href
  {https://doi.org/10.1016/S0006-3495(96)79805-3} {\bibfield  {journal}
  {\bibinfo  {journal} {Biophys. J.}\ }\textbf {\bibinfo {volume} {70}},\
  \bibinfo {pages} {2380} (\bibinfo {year} {1996})}\BibitemShut {NoStop}%
\bibitem [{\citenamefont {B\'{a}lint}\ \emph {et~al.}(2007)\citenamefont
  {B\'{a}lint}, \citenamefont {V\'{e}gh}, \citenamefont {Popescu},
  \citenamefont {Dima}, \citenamefont {Ganea},\ and\ \citenamefont
  {V\'{a}r\'{o}}}]{balint07}%
  \BibitemOpen
  \bibfield  {author} {\bibinfo {author} {\bibfnamefont {Z.}~\bibnamefont
  {B\'{a}lint}}, \bibinfo {author} {\bibfnamefont {G.~A.}\ \bibnamefont
  {V\'{e}gh}}, \bibinfo {author} {\bibfnamefont {A.}~\bibnamefont {Popescu}},
  \bibinfo {author} {\bibfnamefont {M.}~\bibnamefont {Dima}}, \bibinfo {author}
  {\bibfnamefont {C.}~\bibnamefont {Ganea}}, \ and\ \bibinfo {author}
  {\bibfnamefont {G.}~\bibnamefont {V\'{a}r\'{o}}},\ }\href
  {https://dx.doi.org/10.1021/la700666p} {\bibfield  {journal} {\bibinfo
  {journal} {Langmuir}\ }\textbf {\bibinfo {volume} {23}},\ \bibinfo {pages}
  {7225} (\bibinfo {year} {2007})}\BibitemShut {NoStop}%
\bibitem [{\citenamefont {H\"ackl}\ \emph {et~al.}(1998)\citenamefont
  {H\"ackl}, \citenamefont {B\"armann},\ and\ \citenamefont
  {Sackmann}}]{hackl98}%
  \BibitemOpen
  \bibfield  {author} {\bibinfo {author} {\bibfnamefont {W.}~\bibnamefont
  {H\"ackl}}, \bibinfo {author} {\bibfnamefont {M.}~\bibnamefont {B\"armann}},
  \ and\ \bibinfo {author} {\bibfnamefont {E.}~\bibnamefont {Sackmann}},\
  }\href {https://dx.doi.org/10.1103/PhysRevLett.80.1786} {\bibfield  {journal}
  {\bibinfo  {journal} {Phys. Rev. Lett.}\ }\textbf {\bibinfo {volume} {80}},\
  \bibinfo {pages} {1786} (\bibinfo {year} {1998})}\BibitemShut {NoStop}%
\bibitem [{\citenamefont {Bieling}\ \emph {et~al.}(2016)\citenamefont
  {Bieling}, \citenamefont {Li}, \citenamefont {Weichsel}, \citenamefont
  {McGorty}, \citenamefont {Jreij}, \citenamefont {Huang}, \citenamefont
  {Fletcher},\ and\ \citenamefont {Mullins}}]{bieling2016force}%
  \BibitemOpen
  \bibfield  {author} {\bibinfo {author} {\bibfnamefont {P.}~\bibnamefont
  {Bieling}}, \bibinfo {author} {\bibfnamefont {T.-D.}\ \bibnamefont {Li}},
  \bibinfo {author} {\bibfnamefont {J.}~\bibnamefont {Weichsel}}, \bibinfo
  {author} {\bibfnamefont {R.}~\bibnamefont {McGorty}}, \bibinfo {author}
  {\bibfnamefont {P.}~\bibnamefont {Jreij}}, \bibinfo {author} {\bibfnamefont
  {B.}~\bibnamefont {Huang}}, \bibinfo {author} {\bibfnamefont {D.~A.}\
  \bibnamefont {Fletcher}}, \ and\ \bibinfo {author} {\bibfnamefont {R.~D.}\
  \bibnamefont {Mullins}},\ }\href {https://doi.org/10.1016/j.cell.2015.11.057}
  {\bibfield  {journal} {\bibinfo  {journal} {Cell}\ }\textbf {\bibinfo
  {volume} {164}},\ \bibinfo {pages} {115} (\bibinfo {year}
  {2016})}\BibitemShut {NoStop}%
\bibitem [{\citenamefont {Manneville}\ \emph {et~al.}(1999)\citenamefont
  {Manneville}, \citenamefont {Bassereau}, \citenamefont {Levy},\ and\
  \citenamefont {Prost}}]{manneville1999activity}%
  \BibitemOpen
  \bibfield  {author} {\bibinfo {author} {\bibfnamefont {J.-B.}\ \bibnamefont
  {Manneville}}, \bibinfo {author} {\bibfnamefont {P.}~\bibnamefont
  {Bassereau}}, \bibinfo {author} {\bibfnamefont {D.}~\bibnamefont {Levy}}, \
  and\ \bibinfo {author} {\bibfnamefont {J.}~\bibnamefont {Prost}},\ }\href
  {https://doi.org/10.1103/PhysRevLett.82.4356} {\bibfield  {journal} {\bibinfo
   {journal} {Phys. Rev. Lett.}\ }\textbf {\bibinfo {volume} {82}},\ \bibinfo
  {pages} {4356} (\bibinfo {year} {1999})}\BibitemShut {NoStop}%
\bibitem [{\citenamefont {Manneville}\ \emph {et~al.}(2001)\citenamefont
  {Manneville}, \citenamefont {Bassereau}, \citenamefont {Ramaswamy},\ and\
  \citenamefont {Prost}}]{manneville2001active}%
  \BibitemOpen
  \bibfield  {author} {\bibinfo {author} {\bibfnamefont {J.-B.}\ \bibnamefont
  {Manneville}}, \bibinfo {author} {\bibfnamefont {P.}~\bibnamefont
  {Bassereau}}, \bibinfo {author} {\bibfnamefont {S.}~\bibnamefont
  {Ramaswamy}}, \ and\ \bibinfo {author} {\bibfnamefont {J.}~\bibnamefont
  {Prost}},\ }\href {https://doi.org/10.1103/PhysRevE.64.021908} {\bibfield
  {journal} {\bibinfo  {journal} {Phys. Rev. E}\ }\textbf {\bibinfo {volume}
  {64}},\ \bibinfo {pages} {021908} (\bibinfo {year} {2001})}\BibitemShut
  {NoStop}%
\bibitem [{\citenamefont {Chen}(2004)}]{chen2004internal}%
  \BibitemOpen
  \bibfield  {author} {\bibinfo {author} {\bibfnamefont {H.-Y.}\ \bibnamefont
  {Chen}},\ }\href {https://doi.org/10.1103/PhysRevLett.92.168101} {\bibfield
  {journal} {\bibinfo  {journal} {Phys. Rev. Lett.}\ }\textbf {\bibinfo
  {volume} {92}},\ \bibinfo {pages} {168101} (\bibinfo {year}
  {2004})}\BibitemShut {NoStop}%
\bibitem [{\citenamefont {Gov}(2004)}]{gov2004membrane}%
  \BibitemOpen
  \bibfield  {author} {\bibinfo {author} {\bibfnamefont {N.}~\bibnamefont
  {Gov}},\ }\href {https://doi.org/10.1103/PhysRevLett.93.268104} {\bibfield
  {journal} {\bibinfo  {journal} {Phys. Rev. Lett.}\ }\textbf {\bibinfo
  {volume} {93}},\ \bibinfo {pages} {268104} (\bibinfo {year}
  {2004})}\BibitemShut {NoStop}%
\bibitem [{\citenamefont {Lomholt}(2006)}]{lomholt2006fluctuation}%
  \BibitemOpen
  \bibfield  {author} {\bibinfo {author} {\bibfnamefont {M.~A.}\ \bibnamefont
  {Lomholt}},\ }\href {https://doi.org/10.1103/PhysRevE.73.061914} {\bibfield
  {journal} {\bibinfo  {journal} {Phys. Rev. E}\ }\textbf {\bibinfo {volume}
  {73}},\ \bibinfo {pages} {061914} (\bibinfo {year} {2006})}\BibitemShut
  {NoStop}%
\bibitem [{\citenamefont {Lin}\ \emph {et~al.}(2006)\citenamefont {Lin},
  \citenamefont {Gov},\ and\ \citenamefont {Brown}}]{lin2006nonequilibrium}%
  \BibitemOpen
  \bibfield  {author} {\bibinfo {author} {\bibfnamefont {L.~C.-L.}\
  \bibnamefont {Lin}}, \bibinfo {author} {\bibfnamefont {N.}~\bibnamefont
  {Gov}}, \ and\ \bibinfo {author} {\bibfnamefont {F.~L.~H.}\ \bibnamefont
  {Brown}},\ }\href {https://doi.org/10.1063/1.2166383} {\bibfield  {journal}
  {\bibinfo  {journal} {J. Chem. Phys.}\ }\textbf {\bibinfo {volume} {124}},\
  \bibinfo {pages} {074903} (\bibinfo {year} {2006})}\BibitemShut {NoStop}%
\bibitem [{\citenamefont {Loubet}\ \emph {et~al.}(2012)\citenamefont {Loubet},
  \citenamefont {Seifert},\ and\ \citenamefont
  {Lomholt}}]{loubet2012effective}%
  \BibitemOpen
  \bibfield  {author} {\bibinfo {author} {\bibfnamefont {B.}~\bibnamefont
  {Loubet}}, \bibinfo {author} {\bibfnamefont {U.}~\bibnamefont {Seifert}}, \
  and\ \bibinfo {author} {\bibfnamefont {M.~A.}\ \bibnamefont {Lomholt}},\
  }\href {https://doi.org/10.1103/PhysRevE.85.031913} {\bibfield  {journal}
  {\bibinfo  {journal} {Phys. Rev. E}\ }\textbf {\bibinfo {volume} {85}},\
  \bibinfo {pages} {031913} (\bibinfo {year} {2012})}\BibitemShut {NoStop}%
\bibitem [{\citenamefont {Turlier}\ and\ \citenamefont
  {Betz}(2018)}]{Turlier18}%
  \BibitemOpen
  \bibfield  {author} {\bibinfo {author} {\bibfnamefont {H.}~\bibnamefont
  {Turlier}}\ and\ \bibinfo {author} {\bibfnamefont {T.}~\bibnamefont {Betz}},\
  }\enquote {\bibinfo {title} {Fluctuations in active membranes},}\ in\ \href
  {\doibase 10.1007/978-3-030-00630-3_21} {\emph {\bibinfo {booktitle} {Physics
  of Biological Membranes}}},\ \bibinfo {editor} {edited by\ \bibinfo {editor}
  {\bibfnamefont {P.}~\bibnamefont {Bassereau}}\ and\ \bibinfo {editor}
  {\bibfnamefont {P.}~\bibnamefont {Sens}}}\ (\bibinfo  {publisher} {Springer
  International Publishing},\ \bibinfo {address} {Cham},\ \bibinfo {year}
  {2018})\ pp.\ \bibinfo {pages} {581--619}\BibitemShut {NoStop}%
\bibitem [{\citenamefont {Prost}\ and\ \citenamefont
  {Bruinsma}(1996)}]{Prost96}%
  \BibitemOpen
  \bibfield  {author} {\bibinfo {author} {\bibfnamefont {J.}~\bibnamefont
  {Prost}}\ and\ \bibinfo {author} {\bibfnamefont {R.}~\bibnamefont
  {Bruinsma}},\ }\href {https://dx.doi.org/10.1209/epl/i1996-00340-1}
  {\bibfield  {journal} {\bibinfo  {journal} {Europhys. Lett.}\ }\textbf
  {\bibinfo {volume} {33}},\ \bibinfo {pages} {321} (\bibinfo {year}
  {1996})}\BibitemShut {NoStop}%
\bibitem [{\citenamefont {Ramaswamy}\ \emph {et~al.}(2000)\citenamefont
  {Ramaswamy}, \citenamefont {Toner},\ and\ \citenamefont
  {Prost}}]{ramaswamy2000nonequilibrium}%
  \BibitemOpen
  \bibfield  {author} {\bibinfo {author} {\bibfnamefont {S.}~\bibnamefont
  {Ramaswamy}}, \bibinfo {author} {\bibfnamefont {J.}~\bibnamefont {Toner}}, \
  and\ \bibinfo {author} {\bibfnamefont {J.}~\bibnamefont {Prost}},\ }\href
  {https://doi.org/10.1103/PhysRevLett.84.3494} {\bibfield  {journal} {\bibinfo
   {journal} {Phys. Rev. Lett.}\ }\textbf {\bibinfo {volume} {84}},\ \bibinfo
  {pages} {3494} (\bibinfo {year} {2000})}\BibitemShut {NoStop}%
\bibitem [{\citenamefont {Lacoste}\ and\ \citenamefont
  {Lau}(2005)}]{lacoste2005dynamics}%
  \BibitemOpen
  \bibfield  {author} {\bibinfo {author} {\bibfnamefont {D.}~\bibnamefont
  {Lacoste}}\ and\ \bibinfo {author} {\bibfnamefont {A.~W.~C.}\ \bibnamefont
  {Lau}},\ }\href {https://dx.doi.org/10.1209/epl/i2004-10494-8} {\bibfield
  {journal} {\bibinfo  {journal} {Europhys. Lett.}\ }\textbf {\bibinfo {volume}
  {70}},\ \bibinfo {pages} {418} (\bibinfo {year} {2005})}\BibitemShut
  {NoStop}%
\bibitem [{\citenamefont {Ben-Isaac}\ \emph {et~al.}(2011)\citenamefont
  {Ben-Isaac}, \citenamefont {Park}, \citenamefont {Popescu}, \citenamefont
  {Brown}, \citenamefont {Gov},\ and\ \citenamefont
  {Shokef}}]{ben2011effective}%
  \BibitemOpen
  \bibfield  {author} {\bibinfo {author} {\bibfnamefont {E.}~\bibnamefont
  {Ben-Isaac}}, \bibinfo {author} {\bibfnamefont {Y.}~\bibnamefont {Park}},
  \bibinfo {author} {\bibfnamefont {G.}~\bibnamefont {Popescu}}, \bibinfo
  {author} {\bibfnamefont {F.~L.~H.}\ \bibnamefont {Brown}}, \bibinfo {author}
  {\bibfnamefont {N.~S.}\ \bibnamefont {Gov}}, \ and\ \bibinfo {author}
  {\bibfnamefont {Y.}~\bibnamefont {Shokef}},\ }\href {\doibase
  10.1103/PhysRevLett.106.238103} {\bibfield  {journal} {\bibinfo  {journal}
  {Phys. Rev. Lett.}\ }\textbf {\bibinfo {volume} {106}},\ \bibinfo {pages}
  {238103} (\bibinfo {year} {2011})}\BibitemShut {NoStop}%
\bibitem [{\citenamefont {Lin}\ and\ \citenamefont {Brown}(2004)}]{LinBrown04}%
  \BibitemOpen
  \bibfield  {author} {\bibinfo {author} {\bibfnamefont {L.~C.-L.}\
  \bibnamefont {Lin}}\ and\ \bibinfo {author} {\bibfnamefont {F.~L.~H.}\
  \bibnamefont {Brown}},\ }\href {\doibase 10.1103/PhysRevLett.93.256001}
  {\bibfield  {journal} {\bibinfo  {journal} {Phys. Rev. Lett.}\ }\textbf
  {\bibinfo {volume} {93}},\ \bibinfo {pages} {256001} (\bibinfo {year}
  {2004})}\BibitemShut {NoStop}%
\bibitem [{\citenamefont {Gov}\ and\ \citenamefont
  {Gopinathan}(2006)}]{gov2006dynamics}%
  \BibitemOpen
  \bibfield  {author} {\bibinfo {author} {\bibfnamefont {N.~S.}\ \bibnamefont
  {Gov}}\ and\ \bibinfo {author} {\bibfnamefont {A.}~\bibnamefont
  {Gopinathan}},\ }\href {https://doi.org/10.1529/biophysj.105.062224}
  {\bibfield  {journal} {\bibinfo  {journal} {Biophys. J.}\ }\textbf {\bibinfo
  {volume} {90}},\ \bibinfo {pages} {454} (\bibinfo {year} {2006})}\BibitemShut
  {NoStop}%
\bibitem [{\citenamefont {Alert}\ \emph {et~al.}(2015)\citenamefont {Alert},
  \citenamefont {Casademunt}, \citenamefont {Brugu\'{e}s},\ and\ \citenamefont
  {Sens}}]{alert15}%
  \BibitemOpen
  \bibfield  {author} {\bibinfo {author} {\bibfnamefont {R.}~\bibnamefont
  {Alert}}, \bibinfo {author} {\bibfnamefont {J.}~\bibnamefont {Casademunt}},
  \bibinfo {author} {\bibfnamefont {J.}~\bibnamefont {Brugu\'{e}s}}, \ and\
  \bibinfo {author} {\bibfnamefont {P.}~\bibnamefont {Sens}},\ }\href
  {https://doi.org/10.1016/j.bpj.2015.02.027} {\bibfield  {journal} {\bibinfo
  {journal} {Biophys. J.}\ }\textbf {\bibinfo {volume} {108}},\ \bibinfo
  {pages} {1878} (\bibinfo {year} {2015})}\BibitemShut {NoStop}%
\bibitem [{\citenamefont {Brochard}\ and\ \citenamefont
  {Lennon}(1975)}]{brochard75}%
  \BibitemOpen
  \bibfield  {author} {\bibinfo {author} {\bibfnamefont {F.}~\bibnamefont
  {Brochard}}\ and\ \bibinfo {author} {\bibfnamefont {J.}~\bibnamefont
  {Lennon}},\ }\href {https://doi.org/10.1051/jphys:0197500360110103500}
  {\bibfield  {journal} {\bibinfo  {journal} {J. Phys. (France)}\ }\textbf
  {\bibinfo {volume} {36}},\ \bibinfo {pages} {1035} (\bibinfo {year}
  {1975})}\BibitemShut {NoStop}%
\bibitem [{\citenamefont {Gov}\ \emph {et~al.}(2003)\citenamefont {Gov},
  \citenamefont {Zilman},\ and\ \citenamefont {Safran}}]{Gov03}%
  \BibitemOpen
  \bibfield  {author} {\bibinfo {author} {\bibfnamefont {N.}~\bibnamefont
  {Gov}}, \bibinfo {author} {\bibfnamefont {A.~G.}\ \bibnamefont {Zilman}}, \
  and\ \bibinfo {author} {\bibfnamefont {S.}~\bibnamefont {Safran}},\ }\href
  {https://dx.doi.org/10.1103/PhysRevLett.90.228101} {\bibfield  {journal}
  {\bibinfo  {journal} {Phys. Rev. Lett.}\ }\textbf {\bibinfo {volume} {90}},\
  \bibinfo {pages} {228101} (\bibinfo {year} {2003})}\BibitemShut {NoStop}%
\bibitem [{\citenamefont {Fournier}\ \emph {et~al.}(2004)\citenamefont
  {Fournier}, \citenamefont {Lacoste},\ and\ \citenamefont
  {Rapha\"el}}]{fournier04}%
  \BibitemOpen
  \bibfield  {author} {\bibinfo {author} {\bibfnamefont {J.-B.}\ \bibnamefont
  {Fournier}}, \bibinfo {author} {\bibfnamefont {D.}~\bibnamefont {Lacoste}}, \
  and\ \bibinfo {author} {\bibfnamefont {E.}~\bibnamefont {Rapha\"el}},\ }\href
  {https://dx.doi.org/10.1103/PhysRevLett.92.018102} {\bibfield  {journal}
  {\bibinfo  {journal} {Phys. Rev. Lett.}\ }\textbf {\bibinfo {volume} {92}},\
  \bibinfo {pages} {018102} (\bibinfo {year} {2004})}\BibitemShut {NoStop}%
\bibitem [{\citenamefont {Gov}\ and\ \citenamefont
  {Safran}(2005)}]{gov2005red}%
  \BibitemOpen
  \bibfield  {author} {\bibinfo {author} {\bibfnamefont {N.~S.}\ \bibnamefont
  {Gov}}\ and\ \bibinfo {author} {\bibfnamefont {S.~A.}\ \bibnamefont
  {Safran}},\ }\href {https://doi.org/10.1529/biophysj.104.045328} {\bibfield
  {journal} {\bibinfo  {journal} {Biophys. J.}\ }\textbf {\bibinfo {volume}
  {88}},\ \bibinfo {pages} {1859} (\bibinfo {year} {2005})}\BibitemShut
  {NoStop}%
\bibitem [{\citenamefont {Gov}(2007)}]{gov2007active}%
  \BibitemOpen
  \bibfield  {author} {\bibinfo {author} {\bibfnamefont {N.~S.}\ \bibnamefont
  {Gov}},\ }\href {https://doi.org/10.1103/PhysRevE.75.011921} {\bibfield
  {journal} {\bibinfo  {journal} {Phys. Rev. E}\ }\textbf {\bibinfo {volume}
  {75}},\ \bibinfo {pages} {011921} (\bibinfo {year} {2007})}\BibitemShut
  {NoStop}%
\bibitem [{\citenamefont {Turlier}\ \emph {et~al.}(2016)\citenamefont
  {Turlier}, \citenamefont {Fedosov}, \citenamefont {Audoly}, \citenamefont
  {Auth}, \citenamefont {Gov}, \citenamefont {Sykes}, \citenamefont {Joanny},
  \citenamefont {Gompper},\ and\ \citenamefont
  {Betz}}]{turlier2016equilibrium}%
  \BibitemOpen
  \bibfield  {author} {\bibinfo {author} {\bibfnamefont {H.}~\bibnamefont
  {Turlier}}, \bibinfo {author} {\bibfnamefont {D.~A.}\ \bibnamefont
  {Fedosov}}, \bibinfo {author} {\bibfnamefont {B.}~\bibnamefont {Audoly}},
  \bibinfo {author} {\bibfnamefont {T.}~\bibnamefont {Auth}}, \bibinfo {author}
  {\bibfnamefont {N.~S.}\ \bibnamefont {Gov}}, \bibinfo {author} {\bibfnamefont
  {C.}~\bibnamefont {Sykes}}, \bibinfo {author} {\bibfnamefont {J.-F.}\
  \bibnamefont {Joanny}}, \bibinfo {author} {\bibfnamefont {G.}~\bibnamefont
  {Gompper}}, \ and\ \bibinfo {author} {\bibfnamefont {T.}~\bibnamefont
  {Betz}},\ }\href {https://doi.org/10.1038/nphys3621} {\bibfield  {journal}
  {\bibinfo  {journal} {Nat. Phys.}\ }\textbf {\bibinfo {volume} {12}},\
  \bibinfo {pages} {513} (\bibinfo {year} {2016})}\BibitemShut {NoStop}%
\bibitem [{\citenamefont {Paoluzzi}\ \emph {et~al.}(2016)\citenamefont
  {Paoluzzi}, \citenamefont {Di~Leonardo}, \citenamefont {Marchetti},\ and\
  \citenamefont {Angelani}}]{Paoluzzi16}%
  \BibitemOpen
  \bibfield  {author} {\bibinfo {author} {\bibfnamefont {M.}~\bibnamefont
  {Paoluzzi}}, \bibinfo {author} {\bibfnamefont {R.}~\bibnamefont
  {Di~Leonardo}}, \bibinfo {author} {\bibfnamefont {M.~C.}\ \bibnamefont
  {Marchetti}}, \ and\ \bibinfo {author} {\bibfnamefont {L.}~\bibnamefont
  {Angelani}},\ }\href {\doibase 10.1038/srep34146} {\bibfield  {journal}
  {\bibinfo  {journal} {Sci. Rep.}\ }\textbf {\bibinfo {volume} {6}},\ \bibinfo
  {pages} {34146} (\bibinfo {year} {2016})}\BibitemShut {NoStop}%
\bibitem [{\citenamefont {Chen}\ \emph {et~al.}(2017)\citenamefont {Chen},
  \citenamefont {Hua}, \citenamefont {Jiang}, \citenamefont {Zhou},\ and\
  \citenamefont {Zhang}}]{Chen17}%
  \BibitemOpen
  \bibfield  {author} {\bibinfo {author} {\bibfnamefont {J.}~\bibnamefont
  {Chen}}, \bibinfo {author} {\bibfnamefont {Y.}~\bibnamefont {Hua}}, \bibinfo
  {author} {\bibfnamefont {Y.}~\bibnamefont {Jiang}}, \bibinfo {author}
  {\bibfnamefont {X.}~\bibnamefont {Zhou}}, \ and\ \bibinfo {author}
  {\bibfnamefont {L.}~\bibnamefont {Zhang}},\ }\href
  {https://doi.org/10.1038/s41598-017-15095-0} {\bibfield  {journal} {\bibinfo
  {journal} {Sci. Rep.}\ }\textbf {\bibinfo {volume} {7}},\ \bibinfo {pages}
  {15006} (\bibinfo {year} {2017})}\BibitemShut {NoStop}%
\bibitem [{\citenamefont {Wang}\ \emph {et~al.}(2019)\citenamefont {Wang},
  \citenamefont {Guo}, \citenamefont {Tian},\ and\ \citenamefont
  {Chen}}]{Chao19}%
  \BibitemOpen
  \bibfield  {author} {\bibinfo {author} {\bibfnamefont {C.}~\bibnamefont
  {Wang}}, \bibinfo {author} {\bibfnamefont {Y.-K.}\ \bibnamefont {Guo}},
  \bibinfo {author} {\bibfnamefont {W.-D.}\ \bibnamefont {Tian}}, \ and\
  \bibinfo {author} {\bibfnamefont {K.}~\bibnamefont {Chen}},\ }\href
  {https://doi.org/10.1063/1.5078694} {\bibfield  {journal} {\bibinfo
  {journal} {J. Chem. Phys.}\ }\textbf {\bibinfo {volume} {150}},\ \bibinfo
  {pages} {044907} (\bibinfo {year} {2019})}\BibitemShut {NoStop}%
\bibitem [{\citenamefont {Li}\ and\ \citenamefont {ten Wolde}(2019)}]{Li19}%
  \BibitemOpen
  \bibfield  {author} {\bibinfo {author} {\bibfnamefont {Y.}~\bibnamefont
  {Li}}\ and\ \bibinfo {author} {\bibfnamefont {P.~R.}\ \bibnamefont {ten
  Wolde}},\ }\href {\doibase 10.1103/PhysRevLett.123.148003} {\bibfield
  {journal} {\bibinfo  {journal} {Phys. Rev. Lett.}\ }\textbf {\bibinfo
  {volume} {123}},\ \bibinfo {pages} {148003} (\bibinfo {year}
  {2019})}\BibitemShut {NoStop}%
\bibitem [{\citenamefont {Vutukuri}\ \emph {et~al.}(2019)\citenamefont
  {Vutukuri}, \citenamefont {Hoore}, \citenamefont {Abaurrea-Velasco},
  \citenamefont {van Buren}, \citenamefont {Dutto}, \citenamefont {Auth},
  \citenamefont {Fedosov}, \citenamefont {Gompper},\ and\ \citenamefont
  {Vermant}}]{vutukuri2019sculpting}%
  \BibitemOpen
  \bibfield  {author} {\bibinfo {author} {\bibfnamefont {H.~R.}\ \bibnamefont
  {Vutukuri}}, \bibinfo {author} {\bibfnamefont {M.}~\bibnamefont {Hoore}},
  \bibinfo {author} {\bibfnamefont {C.}~\bibnamefont {Abaurrea-Velasco}},
  \bibinfo {author} {\bibfnamefont {L.}~\bibnamefont {van Buren}}, \bibinfo
  {author} {\bibfnamefont {A.}~\bibnamefont {Dutto}}, \bibinfo {author}
  {\bibfnamefont {T.}~\bibnamefont {Auth}}, \bibinfo {author} {\bibfnamefont
  {D.~A.}\ \bibnamefont {Fedosov}}, \bibinfo {author} {\bibfnamefont
  {G.}~\bibnamefont {Gompper}}, \ and\ \bibinfo {author} {\bibfnamefont
  {J.}~\bibnamefont {Vermant}},\ }\href@noop {} {\  (\bibinfo {year} {2019})},\
  \Eprint {http://arxiv.org/abs/1911.02381} {arXiv:1911.02381} \BibitemShut
  {NoStop}%
\bibitem [{Sup()}]{Supplemental}%
  \BibitemOpen
  \href@noop {} {}\bibinfo {note} {See Supplemental Material (SM) below, which
  includes experimental videos of passive and active membranes fluctuating, as
  well as a description of experimental and numerical protocols.}\BibitemShut
  {Stop}%
\bibitem [{\citenamefont {Angelova}\ and\ \citenamefont
  {Dimitrov}(1986)}]{Angelo75}%
  \BibitemOpen
  \bibfield  {author} {\bibinfo {author} {\bibfnamefont {M.~I.}\ \bibnamefont
  {Angelova}}\ and\ \bibinfo {author} {\bibfnamefont {D.~S.}\ \bibnamefont
  {Dimitrov}},\ }\href {\doibase 10.1039/DC9868100303} {\bibfield  {journal}
  {\bibinfo  {journal} {Faraday Discuss. Chem. Soc.}\ }\textbf {\bibinfo
  {volume} {81}},\ \bibinfo {pages} {303} (\bibinfo {year} {1986})}\BibitemShut
  {NoStop}%
\bibitem [{\citenamefont {Kuribayashi}\ \emph {et~al.}(2006)\citenamefont
  {Kuribayashi}, \citenamefont {Tresset}, \citenamefont {Coquet}, \citenamefont
  {Fujita},\ and\ \citenamefont {Takeuchi}}]{Kuribayashi06}%
  \BibitemOpen
  \bibfield  {author} {\bibinfo {author} {\bibfnamefont {K.}~\bibnamefont
  {Kuribayashi}}, \bibinfo {author} {\bibfnamefont {G.}~\bibnamefont
  {Tresset}}, \bibinfo {author} {\bibfnamefont {P.}~\bibnamefont {Coquet}},
  \bibinfo {author} {\bibfnamefont {H.}~\bibnamefont {Fujita}}, \ and\ \bibinfo
  {author} {\bibfnamefont {S.}~\bibnamefont {Takeuchi}},\ }\href {\doibase
  10.1088/0957-0233/17/12/S01} {\bibfield  {journal} {\bibinfo  {journal}
  {Meas. Sci. Technol.}\ }\textbf {\bibinfo {volume} {17}},\ \bibinfo {pages}
  {3121} (\bibinfo {year} {2006})}\BibitemShut {NoStop}%
\bibitem [{Note1()}]{Note1}%
  \BibitemOpen
  \bibinfo {note} {Another possibility is that electroformation temporarily
  weakens the bacteria, and it takes them time to recover.}\BibitemShut {Stop}%
\bibitem [{Note2()}]{Note2}%
  \BibitemOpen
  \bibinfo {note} {Bacterial division occurs on a time scale of $\sim \protect
  \tmspace -\thinmuskip {.1667em} 30$--60 min, and so does not affect our
  measurements}\BibitemShut {NoStop}%
\bibitem [{\citenamefont {P{\'e}cr{\'e}aux}\ \emph {et~al.}(2004)\citenamefont
  {P{\'e}cr{\'e}aux}, \citenamefont {D{\"o}bereiner}, \citenamefont {Prost},
  \citenamefont {Joanny},\ and\ \citenamefont {Bassereau}}]{Pecreaux04}%
  \BibitemOpen
  \bibfield  {author} {\bibinfo {author} {\bibfnamefont {J.}~\bibnamefont
  {P{\'e}cr{\'e}aux}}, \bibinfo {author} {\bibfnamefont {H.-G.}\ \bibnamefont
  {D{\"o}bereiner}}, \bibinfo {author} {\bibfnamefont {J.}~\bibnamefont
  {Prost}}, \bibinfo {author} {\bibfnamefont {J.-F.}\ \bibnamefont {Joanny}}, \
  and\ \bibinfo {author} {\bibfnamefont {P.}~\bibnamefont {Bassereau}},\ }\href
  {\doibase 10.1140/epje/i2004-10001-9} {\bibfield  {journal} {\bibinfo
  {journal} {Eur. Phys. J. E}\ }\textbf {\bibinfo {volume} {13}},\ \bibinfo
  {pages} {277} (\bibinfo {year} {2004})}\BibitemShut {NoStop}%
\bibitem [{\citenamefont {Gracià}\ \emph {et~al.}(2010)\citenamefont
  {Gracià}, \citenamefont {Bezlyepkina}, \citenamefont {Knorr}, \citenamefont
  {Lipowsky},\ and\ \citenamefont {Dimova}}]{Gracia10}%
  \BibitemOpen
  \bibfield  {author} {\bibinfo {author} {\bibfnamefont {R.~S.}\ \bibnamefont
  {Gracià}}, \bibinfo {author} {\bibfnamefont {N.}~\bibnamefont
  {Bezlyepkina}}, \bibinfo {author} {\bibfnamefont {R.~L.}\ \bibnamefont
  {Knorr}}, \bibinfo {author} {\bibfnamefont {R.}~\bibnamefont {Lipowsky}}, \
  and\ \bibinfo {author} {\bibfnamefont {R.}~\bibnamefont {Dimova}},\ }\href
  {\doibase 10.1039/B920629A} {\bibfield  {journal} {\bibinfo  {journal} {Soft
  Matter}\ }\textbf {\bibinfo {volume} {6}},\ \bibinfo {pages} {1472} (\bibinfo
  {year} {2010})}\BibitemShut {NoStop}%
\bibitem [{\citenamefont {M{\'e}l{\'e}ard}\ \emph {et~al.}(2011)\citenamefont
  {M{\'e}l{\'e}ard}, \citenamefont {Pott}, \citenamefont {Bouvrais},\ and\
  \citenamefont {Ipsen}}]{Meleard11}%
  \BibitemOpen
  \bibfield  {author} {\bibinfo {author} {\bibfnamefont {P.}~\bibnamefont
  {M{\'e}l{\'e}ard}}, \bibinfo {author} {\bibfnamefont {T.}~\bibnamefont
  {Pott}}, \bibinfo {author} {\bibfnamefont {H.}~\bibnamefont {Bouvrais}}, \
  and\ \bibinfo {author} {\bibfnamefont {J.~H.}\ \bibnamefont {Ipsen}},\ }\href
  {\doibase 10.1140/epje/i2011-11116-6} {\bibfield  {journal} {\bibinfo
  {journal} {Eur. Phys. J. E}\ }\textbf {\bibinfo {volume} {34}},\ \bibinfo
  {pages} {116} (\bibinfo {year} {2011})}\BibitemShut {NoStop}%
\bibitem [{Note3()}]{Note3}%
  \BibitemOpen
  \bibinfo {note} {We have verified from active particle simulations that small
  errors in detecting the vesicle center of mass do not significantly affect
  the results for modes $m \geq 2$.}\BibitemShut {Stop}%
\bibitem [{\citenamefont {Monge}(1807)}]{monge}%
  \BibitemOpen
  \bibfield  {author} {\bibinfo {author} {\bibfnamefont {G.}~\bibnamefont
  {Monge}},\ }\href@noop {} {\emph {\bibinfo {title} {Application de l'analyse
  \`{a} la G\'{e}om\'{e}trie}}}\ (\bibinfo  {publisher} {Bernard},\ \bibinfo
  {year} {1807})\BibitemShut {NoStop}%
\bibitem [{\citenamefont {Sahu}\ \emph {et~al.}(2019)\citenamefont {Sahu},
  \citenamefont {Glisman}, \citenamefont {Tchoufag},\ and\ \citenamefont
  {Mandadapu}}]{Sahu19}%
  \BibitemOpen
  \bibfield  {author} {\bibinfo {author} {\bibfnamefont {A.}~\bibnamefont
  {Sahu}}, \bibinfo {author} {\bibfnamefont {A.}~\bibnamefont {Glisman}},
  \bibinfo {author} {\bibfnamefont {J.}~\bibnamefont {Tchoufag}}, \ and\
  \bibinfo {author} {\bibfnamefont {K.~K.}\ \bibnamefont {Mandadapu}},\
  }\href@noop {} {\  (\bibinfo {year} {2019})},\ \Eprint
  {http://arxiv.org/abs/1910.10693} {arXiv:1910.10693} \BibitemShut {NoStop}%
\bibitem [{\citenamefont {Canham}(1970)}]{Canham1970}%
  \BibitemOpen
  \bibfield  {author} {\bibinfo {author} {\bibfnamefont {P.~B.}\ \bibnamefont
  {Canham}},\ }\href {https://doi.org/10.1016/S0022-5193(70)80032-7} {\bibfield
   {journal} {\bibinfo  {journal} {J. Theor. Biol.}\ }\textbf {\bibinfo
  {volume} {26}},\ \bibinfo {pages} {61} (\bibinfo {year} {1970})}\BibitemShut
  {NoStop}%
\bibitem [{\citenamefont {Helfrich}(1973)}]{Helfrich73}%
  \BibitemOpen
  \bibfield  {author} {\bibinfo {author} {\bibfnamefont {W.}~\bibnamefont
  {Helfrich}},\ }\href {\doibase 10.1515/znc-1973-11-1209} {\bibfield
  {journal} {\bibinfo  {journal} {Z. Naturforsch.}\ }\textbf {\bibinfo {volume}
  {28C}},\ \bibinfo {pages} {693} (\bibinfo {year} {1973})}\BibitemShut
  {NoStop}%
\bibitem [{\citenamefont {Evans}(1974)}]{Evans74}%
  \BibitemOpen
  \bibfield  {author} {\bibinfo {author} {\bibfnamefont {E.~A.}\ \bibnamefont
  {Evans}},\ }\href {https://doi.org/10.1016/S0006-3495(74)85959-X} {\bibfield
  {journal} {\bibinfo  {journal} {Biophys. J.}\ }\textbf {\bibinfo {volume}
  {14}},\ \bibinfo {pages} {923} (\bibinfo {year} {1974})}\BibitemShut
  {NoStop}%
\bibitem [{\citenamefont {Sapp}\ and\ \citenamefont {Maibaum}(2016)}]{Sapp16}%
  \BibitemOpen
  \bibfield  {author} {\bibinfo {author} {\bibfnamefont {K.}~\bibnamefont
  {Sapp}}\ and\ \bibinfo {author} {\bibfnamefont {L.}~\bibnamefont {Maibaum}},\
  }\href {\doibase 10.1103/PhysRevE.94.052414} {\bibfield  {journal} {\bibinfo
  {journal} {Phys. Rev. E}\ }\textbf {\bibinfo {volume} {94}},\ \bibinfo
  {pages} {052414} (\bibinfo {year} {2016})}\BibitemShut {NoStop}%
\bibitem [{\citenamefont {Drescher}\ \emph {et~al.}(2011)\citenamefont
  {Drescher}, \citenamefont {Dunkel}, \citenamefont {Cisneros}, \citenamefont
  {Ganguly},\ and\ \citenamefont {Goldstein}}]{drescher2011fluid}%
  \BibitemOpen
  \bibfield  {author} {\bibinfo {author} {\bibfnamefont {K.}~\bibnamefont
  {Drescher}}, \bibinfo {author} {\bibfnamefont {J.}~\bibnamefont {Dunkel}},
  \bibinfo {author} {\bibfnamefont {L.~H.}\ \bibnamefont {Cisneros}}, \bibinfo
  {author} {\bibfnamefont {S.}~\bibnamefont {Ganguly}}, \ and\ \bibinfo
  {author} {\bibfnamefont {R.~E.}\ \bibnamefont {Goldstein}},\ }\href {\doibase
  10.1073/pnas.1019079108} {\bibfield  {journal} {\bibinfo  {journal} {Proc.
  Natl. Acad. Sci. U.S.A.}\ }\textbf {\bibinfo {volume} {108}},\ \bibinfo
  {pages} {10940} (\bibinfo {year} {2011})}\BibitemShut {NoStop}%
\bibitem [{Note4()}]{Note4}%
  \BibitemOpen
  \bibinfo {note} {Our code is publicly available at \protect \texttt {\protect
  \href {https://github.com/mandadapu-group/active-contact}{https://github.com/
  mandadapu-group/active-contact}}}\BibitemShut {NoStop}%
\bibitem [{\citenamefont {Yan}\ and\ \citenamefont
  {Brady}(2015)}]{yan2015force}%
  \BibitemOpen
  \bibfield  {author} {\bibinfo {author} {\bibfnamefont {W.}~\bibnamefont
  {Yan}}\ and\ \bibinfo {author} {\bibfnamefont {J.~F.}\ \bibnamefont
  {Brady}},\ }\href {\doibase 10.1017/jfm.2015.621} {\bibfield  {journal}
  {\bibinfo  {journal} {J. Fluid Mech.}\ }\textbf {\bibinfo {volume} {785}},\
  \bibinfo {pages} {R1} (\bibinfo {year} {2015})}\BibitemShut {NoStop}%
\bibitem [{\citenamefont {Nikola}\ \emph {et~al.}(2016)\citenamefont {Nikola},
  \citenamefont {Solon}, \citenamefont {Kafri}, \citenamefont {Kardar},
  \citenamefont {Tailleur},\ and\ \citenamefont
  {Voituriez}}]{nikola2016active}%
  \BibitemOpen
  \bibfield  {author} {\bibinfo {author} {\bibfnamefont {N.}~\bibnamefont
  {Nikola}}, \bibinfo {author} {\bibfnamefont {A.~P.}\ \bibnamefont {Solon}},
  \bibinfo {author} {\bibfnamefont {Y.}~\bibnamefont {Kafri}}, \bibinfo
  {author} {\bibfnamefont {M.}~\bibnamefont {Kardar}}, \bibinfo {author}
  {\bibfnamefont {J.}~\bibnamefont {Tailleur}}, \ and\ \bibinfo {author}
  {\bibfnamefont {R.}~\bibnamefont {Voituriez}},\ }\href
  {10.1103/PhysRevLett.117.098001} {\bibfield  {journal} {\bibinfo  {journal}
  {Phys. Rev. Lett.}\ }\textbf {\bibinfo {volume} {117}},\ \bibinfo {pages}
  {098001} (\bibinfo {year} {2016})}\BibitemShut {NoStop}%
\bibitem [{\citenamefont {Takatori}\ \emph {et~al.}(2016)\citenamefont
  {Takatori}, \citenamefont {De~Dier}, \citenamefont {Vermant},\ and\
  \citenamefont {Brady}}]{Takatori16a}%
  \BibitemOpen
  \bibfield  {author} {\bibinfo {author} {\bibfnamefont {S.~C.}\ \bibnamefont
  {Takatori}}, \bibinfo {author} {\bibfnamefont {R.}~\bibnamefont {De~Dier}},
  \bibinfo {author} {\bibfnamefont {J.}~\bibnamefont {Vermant}}, \ and\
  \bibinfo {author} {\bibfnamefont {J.~F.}\ \bibnamefont {Brady}},\ }\href
  {\doibase 10.1038/ncomms10694} {\bibfield  {journal} {\bibinfo  {journal}
  {Nat. Commun.}\ }\textbf {\bibinfo {volume} {7}},\ \bibinfo {pages} {10694}
  (\bibinfo {year} {2016})}\BibitemShut {NoStop}%
\bibitem [{\citenamefont {Baraban}\ \emph {et~al.}(2013)\citenamefont
  {Baraban}, \citenamefont {Makarov}, \citenamefont {Schmidt}, \citenamefont
  {Cuniberti}, \citenamefont {Leiderer},\ and\ \citenamefont
  {Erbe}}]{Baraban13}%
  \BibitemOpen
  \bibfield  {author} {\bibinfo {author} {\bibfnamefont {L.}~\bibnamefont
  {Baraban}}, \bibinfo {author} {\bibfnamefont {D.}~\bibnamefont {Makarov}},
  \bibinfo {author} {\bibfnamefont {O.~G.}\ \bibnamefont {Schmidt}}, \bibinfo
  {author} {\bibfnamefont {G.}~\bibnamefont {Cuniberti}}, \bibinfo {author}
  {\bibfnamefont {P.}~\bibnamefont {Leiderer}}, \ and\ \bibinfo {author}
  {\bibfnamefont {A.}~\bibnamefont {Erbe}},\ }\href {\doibase
  10.1039/C2NR32662K} {\bibfield  {journal} {\bibinfo  {journal} {Nanoscale}\
  }\textbf {\bibinfo {volume} {5}},\ \bibinfo {pages} {1332} (\bibinfo {year}
  {2013})}\BibitemShut {NoStop}%
\bibitem [{\citenamefont {Friedrich}\ \emph {et~al.}(2012)\citenamefont
  {Friedrich}, \citenamefont {Hagedorn}, \citenamefont {Soldati-Favre},\ and\
  \citenamefont {Soldati}}]{Friedrich12}%
  \BibitemOpen
  \bibfield  {author} {\bibinfo {author} {\bibfnamefont {N.}~\bibnamefont
  {Friedrich}}, \bibinfo {author} {\bibfnamefont {M.}~\bibnamefont {Hagedorn}},
  \bibinfo {author} {\bibfnamefont {D.}~\bibnamefont {Soldati-Favre}}, \ and\
  \bibinfo {author} {\bibfnamefont {T.}~\bibnamefont {Soldati}},\ }\href
  {\doibase 10.1128/MMBR.00024-12} {\bibfield  {journal} {\bibinfo  {journal}
  {Microbiol. Mol. Biol. Rev.}\ }\textbf {\bibinfo {volume} {76}},\ \bibinfo
  {pages} {707} (\bibinfo {year} {2012})}\BibitemShut {NoStop}%
\bibitem [{\citenamefont {Pizarro-Cerd\'{a}}\ \emph {et~al.}(2016)\citenamefont
  {Pizarro-Cerd\'{a}}, \citenamefont {Charbit}, \citenamefont {Enninga},
  \citenamefont {Lafont},\ and\ \citenamefont {Cossart}}]{Pizarro16}%
  \BibitemOpen
  \bibfield  {author} {\bibinfo {author} {\bibfnamefont {J.}~\bibnamefont
  {Pizarro-Cerd\'{a}}}, \bibinfo {author} {\bibfnamefont {A.}~\bibnamefont
  {Charbit}}, \bibinfo {author} {\bibfnamefont {J.}~\bibnamefont {Enninga}},
  \bibinfo {author} {\bibfnamefont {F.}~\bibnamefont {Lafont}}, \ and\ \bibinfo
  {author} {\bibfnamefont {P.}~\bibnamefont {Cossart}},\ }\href {\doibase
  https://doi.org/10.1016/j.semcdb.2016.07.019} {\bibfield  {journal} {\bibinfo
   {journal} {Semin. Cell Dev. Biol.}\ }\textbf {\bibinfo {volume} {60}},\
  \bibinfo {pages} {155} (\bibinfo {year} {2016})}\BibitemShut {NoStop}%
\bibitem [{\citenamefont {Fo\v{s}nari\v{c}}\ \emph {et~al.}(2019)\citenamefont
  {Fo\v{s}nari\v{c}}, \citenamefont {Peni\v{c}}, \citenamefont {Igli{\v{c}}},
  \citenamefont {Kralj-Igli\v{c}}, \citenamefont {Drab},\ and\ \citenamefont
  {Gov}}]{fosnaric2019theoretical}%
  \BibitemOpen
  \bibfield  {author} {\bibinfo {author} {\bibfnamefont {M.}~\bibnamefont
  {Fo\v{s}nari\v{c}}}, \bibinfo {author} {\bibfnamefont {S.}~\bibnamefont
  {Peni\v{c}}}, \bibinfo {author} {\bibfnamefont {A.}~\bibnamefont
  {Igli{\v{c}}}}, \bibinfo {author} {\bibfnamefont {V.}~\bibnamefont
  {Kralj-Igli\v{c}}}, \bibinfo {author} {\bibfnamefont {M.}~\bibnamefont
  {Drab}}, \ and\ \bibinfo {author} {\bibfnamefont {N.~S.}\ \bibnamefont
  {Gov}},\ }\href {\doibase 10.1039/C8SM02356E} {\bibfield  {journal} {\bibinfo
   {journal} {Soft Matter}\ }\textbf {\bibinfo {volume} {15}},\ \bibinfo
  {pages} {5319} (\bibinfo {year} {2019})}\BibitemShut {NoStop}%
\bibitem [{\citenamefont {Sahu}\ \emph {et~al.}(2017)\citenamefont {Sahu},
  \citenamefont {Sauer},\ and\ \citenamefont {Mandadapu}}]{Sahu17}%
  \BibitemOpen
  \bibfield  {author} {\bibinfo {author} {\bibfnamefont {A.}~\bibnamefont
  {Sahu}}, \bibinfo {author} {\bibfnamefont {R.~A.}\ \bibnamefont {Sauer}}, \
  and\ \bibinfo {author} {\bibfnamefont {K.~K.}\ \bibnamefont {Mandadapu}},\
  }\href {\doibase 10.1103/PhysRevE.96.042409} {\bibfield  {journal} {\bibinfo
  {journal} {Phys. Rev. E}\ }\textbf {\bibinfo {volume} {96}},\ \bibinfo
  {pages} {042409} (\bibinfo {year} {2017})}\BibitemShut {NoStop}%
\bibitem [{\citenamefont {Sahu}\ \emph {et~al.}(2020)\citenamefont {Sahu},
  \citenamefont {Omar}, \citenamefont {Sauer},\ and\ \citenamefont
  {Mandadapu}}]{Sahu18}%
  \BibitemOpen
  \bibfield  {author} {\bibinfo {author} {\bibfnamefont {A.}~\bibnamefont
  {Sahu}}, \bibinfo {author} {\bibfnamefont {Y.~A.~D.}\ \bibnamefont {Omar}},
  \bibinfo {author} {\bibfnamefont {R.~A.}\ \bibnamefont {Sauer}}, \ and\
  \bibinfo {author} {\bibfnamefont {K.~K.}\ \bibnamefont {Mandadapu}},\ }\href
  {\doibase 10.1016/j.jcp.2020.109253} {\bibfield  {journal} {\bibinfo
  {journal} {J. Comp. Phys.}\ }\textbf {\bibinfo {volume} {407}},\ \bibinfo
  {pages} {109253} (\bibinfo {year} {2020})}\BibitemShut {NoStop}%
\bibitem [{\citenamefont {Tsai}\ \emph {et~al.}(2011)\citenamefont {Tsai},
  \citenamefont {Stuhrmann},\ and\ \citenamefont {Koenderink}}]{sm-Tsai11}%
  \BibitemOpen
  \bibfield  {author} {\bibinfo {author} {\bibfnamefont {F.-C.}\ \bibnamefont
  {Tsai}}, \bibinfo {author} {\bibfnamefont {B.}~\bibnamefont {Stuhrmann}}, \
  and\ \bibinfo {author} {\bibfnamefont {G.~H.}\ \bibnamefont {Koenderink}},\
  }\href {\doibase 10.1021/la201604z} {\bibfield  {journal} {\bibinfo
  {journal} {Langmuir}\ }\textbf {\bibinfo {volume} {27}},\ \bibinfo {pages}
  {10061} (\bibinfo {year} {2011})}\BibitemShut {NoStop}%
\bibitem [{\citenamefont {Baird}(1994)}]{sm-Baird94}%
  \BibitemOpen
  \bibfield  {author} {\bibinfo {author} {\bibfnamefont {D.}~\bibnamefont
  {Baird}},\ }\href@noop {} {\emph {\bibinfo {title} {Experimentation: An
  Introduction to Measurement Theory and Experiment Design}}}\ (\bibinfo
  {publisher} {Benjamin Cummings},\ \bibinfo {address} {3rd Ed.},\ \bibinfo
  {year} {1994})\BibitemShut {NoStop}%
\bibitem [{\citenamefont {M{\'{e}}l{\'{e}}ard}\ \emph
  {et~al.}(1992)\citenamefont {M{\'{e}}l{\'{e}}ard}, \citenamefont {Faucon},
  \citenamefont {Mitov},\ and\ \citenamefont {Bothorel}}]{sm-Meleard92}%
  \BibitemOpen
  \bibfield  {author} {\bibinfo {author} {\bibfnamefont {P.}~\bibnamefont
  {M{\'{e}}l{\'{e}}ard}}, \bibinfo {author} {\bibfnamefont {J.}~\bibnamefont
  {Faucon}}, \bibinfo {author} {\bibfnamefont {M.}~\bibnamefont {Mitov}}, \
  and\ \bibinfo {author} {\bibfnamefont {P.}~\bibnamefont {Bothorel}},\ }\href
  {\doibase 10.1209/0295-5075/19/4/004} {\bibfield  {journal} {\bibinfo
  {journal} {Europhys. Lett.}\ }\textbf {\bibinfo {volume} {19}},\ \bibinfo
  {pages} {267} (\bibinfo {year} {1992})}\BibitemShut {NoStop}%
\bibitem [{\citenamefont {Leal}(2007)}]{sm-Leal}%
  \BibitemOpen
  \bibfield  {author} {\bibinfo {author} {\bibfnamefont {L.~G.}\ \bibnamefont
  {Leal}},\ }\href {\doibase 10.1017/CBO9780511800245} {\emph {\bibinfo {title}
  {Advanced Transport Phenomena: Fluid Mechanics and Convective Transport
  Processes}}},\ Cambridge Series in Chemical Engineering\ (\bibinfo
  {publisher} {Cambridge University Press},\ \bibinfo {year}
  {2007})\BibitemShut {NoStop}%
\end{thebibliography}

\begin{thebibliography}{15}%
\makeatletter
\providecommand \@ifxundefined [1]{%
 \@ifx{#1\undefined}
}%
\providecommand \@ifnum [1]{%
 \ifnum #1\expandafter \@firstoftwo
 \else \expandafter \@secondoftwo
 \fi
}%
\providecommand \@ifx [1]{%
 \ifx #1\expandafter \@firstoftwo
 \else \expandafter \@secondoftwo
 \fi
}%
\providecommand \natexlab [1]{#1}%
\providecommand \enquote  [1]{``#1''}%
\providecommand \bibnamefont  [1]{#1}%
\providecommand \bibfnamefont [1]{#1}%
\providecommand \citenamefont [1]{#1}%
\providecommand \href@noop [0]{\@secondoftwo}%
\providecommand \href [0]{\begingroup \@sanitize@url \@href}%
\providecommand \@href[1]{\@@startlink{#1}\@@href}%
\providecommand \@@href[1]{\endgroup#1\@@endlink}%
\providecommand \@sanitize@url [0]{\catcode `\\12\catcode `\$12\catcode
  `\&12\catcode `\#12\catcode `\^12\catcode `\_12\catcode `\%12\relax}%
\providecommand \@@startlink[1]{}%
\providecommand \@@endlink[0]{}%
\providecommand \url  [0]{\begingroup\@sanitize@url \@url }%
\providecommand \@url [1]{\endgroup\@href {#1}{\urlprefix }}%
\providecommand \urlprefix  [0]{URL }%
\providecommand \Eprint [0]{\href }%
\providecommand \doibase [0]{http://dx.doi.org/}%
\providecommand \selectlanguage [0]{\@gobble}%
\providecommand \bibinfo  [0]{\@secondoftwo}%
\providecommand \bibfield  [0]{\@secondoftwo}%
\providecommand \translation [1]{[#1]}%
\providecommand \BibitemOpen [0]{}%
\providecommand \bibitemStop [0]{}%
\providecommand \bibitemNoStop [0]{.\EOS\space}%
\providecommand \EOS [0]{\spacefactor3000\relax}%
\providecommand \BibitemShut  [1]{\csname bibitem#1\endcsname}%
\let\auto@bib@innerbib\@empty
\bibitem [{\citenamefont {P{\'e}cr{\'e}aux}\ \emph {et~al.}(2004)\citenamefont
  {P{\'e}cr{\'e}aux}, \citenamefont {D{\"o}bereiner}, \citenamefont {Prost},
  \citenamefont {Joanny},\ and\ \citenamefont {Bassereau}}]{sm-Pecreaux04}%
  \BibitemOpen
  \bibfield  {author} {\bibinfo {author} {\bibfnamefont {J.}~\bibnamefont
  {P{\'e}cr{\'e}aux}}, \bibinfo {author} {\bibfnamefont {H.-G.}\ \bibnamefont
  {D{\"o}bereiner}}, \bibinfo {author} {\bibfnamefont {J.}~\bibnamefont
  {Prost}}, \bibinfo {author} {\bibfnamefont {J.-F.}\ \bibnamefont {Joanny}}, \
  and\ \bibinfo {author} {\bibfnamefont {P.}~\bibnamefont {Bassereau}},\ }\href
  {\doibase 10.1140/epje/i2004-10001-9} {\bibfield  {journal} {\bibinfo
  {journal} {Eur. Phys. J. E}\ }\textbf {\bibinfo {volume} {13}},\ \bibinfo
  {pages} {277} (\bibinfo {year} {2004})}\BibitemShut {NoStop}%
\bibitem [{\citenamefont {Tsai}\ \emph {et~al.}(2011)\citenamefont {Tsai},
  \citenamefont {Stuhrmann},\ and\ \citenamefont {Koenderink}}]{sm-Tsai11}%
  \BibitemOpen
  \bibfield  {author} {\bibinfo {author} {\bibfnamefont {F.-C.}\ \bibnamefont
  {Tsai}}, \bibinfo {author} {\bibfnamefont {B.}~\bibnamefont {Stuhrmann}}, \
  and\ \bibinfo {author} {\bibfnamefont {G.~H.}\ \bibnamefont {Koenderink}},\
  }\href {\doibase 10.1021/la201604z} {\bibfield  {journal} {\bibinfo
  {journal} {Langmuir}\ }\textbf {\bibinfo {volume} {27}},\ \bibinfo {pages}
  {10061} (\bibinfo {year} {2011})}\BibitemShut {NoStop}%
\bibitem [{\citenamefont {Baird}(1994)}]{sm-Baird94}%
  \BibitemOpen
  \bibfield  {author} {\bibinfo {author} {\bibfnamefont {D.}~\bibnamefont
  {Baird}},\ }\href@noop {} {\emph {\bibinfo {title} {Experimentation: An
  Introduction to Measurement Theory and Experiment Design}}}\ (\bibinfo
  {publisher} {Benjamin Cummings},\ \bibinfo {address} {3rd Ed.},\ \bibinfo
  {year} {1994})\BibitemShut {NoStop}%
\bibitem [{\citenamefont {M{\'{e}}l{\'{e}}ard}\ \emph
  {et~al.}(1992)\citenamefont {M{\'{e}}l{\'{e}}ard}, \citenamefont {Faucon},
  \citenamefont {Mitov},\ and\ \citenamefont {Bothorel}}]{sm-Meleard92}%
  \BibitemOpen
  \bibfield  {author} {\bibinfo {author} {\bibfnamefont {P.}~\bibnamefont
  {M{\'{e}}l{\'{e}}ard}}, \bibinfo {author} {\bibfnamefont {J.}~\bibnamefont
  {Faucon}}, \bibinfo {author} {\bibfnamefont {M.}~\bibnamefont {Mitov}}, \
  and\ \bibinfo {author} {\bibfnamefont {P.}~\bibnamefont {Bothorel}},\ }\href
  {\doibase 10.1209/0295-5075/19/4/004} {\bibfield  {journal} {\bibinfo
  {journal} {Europhys. Lett.}\ }\textbf {\bibinfo {volume} {19}},\ \bibinfo
  {pages} {267} (\bibinfo {year} {1992})}\BibitemShut {NoStop}%
\bibitem [{\citenamefont {M{\'e}l{\'e}ard}\ \emph {et~al.}(2011)\citenamefont
  {M{\'e}l{\'e}ard}, \citenamefont {Pott}, \citenamefont {Bouvrais},\ and\
  \citenamefont {Ipsen}}]{sm-Meleard11}%
  \BibitemOpen
  \bibfield  {author} {\bibinfo {author} {\bibfnamefont {P.}~\bibnamefont
  {M{\'e}l{\'e}ard}}, \bibinfo {author} {\bibfnamefont {T.}~\bibnamefont
  {Pott}}, \bibinfo {author} {\bibfnamefont {H.}~\bibnamefont {Bouvrais}}, \
  and\ \bibinfo {author} {\bibfnamefont {J.~H.}\ \bibnamefont {Ipsen}},\ }\href
  {\doibase 10.1140/epje/i2011-11116-6} {\bibfield  {journal} {\bibinfo
  {journal} {Eur. Phys. J. E}\ }\textbf {\bibinfo {volume} {34}},\ \bibinfo
  {pages} {116} (\bibinfo {year} {2011})}\BibitemShut {NoStop}%
\bibitem [{\citenamefont {Sapp}\ and\ \citenamefont
  {Maibaum}(2016)}]{sm-Sapp16}%
  \BibitemOpen
  \bibfield  {author} {\bibinfo {author} {\bibfnamefont {K.}~\bibnamefont
  {Sapp}}\ and\ \bibinfo {author} {\bibfnamefont {L.}~\bibnamefont {Maibaum}},\
  }\href {\doibase 10.1103/PhysRevE.94.052414} {\bibfield  {journal} {\bibinfo
  {journal} {Phys. Rev. E}\ }\textbf {\bibinfo {volume} {94}},\ \bibinfo
  {pages} {052414} (\bibinfo {year} {2016})}\BibitemShut {NoStop}%
\bibitem [{\citenamefont {Lin}\ and\ \citenamefont
  {Brown}(2004)}]{sm-LinBrown04}%
  \BibitemOpen
  \bibfield  {author} {\bibinfo {author} {\bibfnamefont {L.~C.-L.}\
  \bibnamefont {Lin}}\ and\ \bibinfo {author} {\bibfnamefont {F.~L.~H.}\
  \bibnamefont {Brown}},\ }\href {\doibase 10.1103/PhysRevLett.93.256001}
  {\bibfield  {journal} {\bibinfo  {journal} {Phys. Rev. Lett.}\ }\textbf
  {\bibinfo {volume} {93}},\ \bibinfo {pages} {256001} (\bibinfo {year}
  {2004})}\BibitemShut {NoStop}%
\bibitem [{\citenamefont {Canham}(1970)}]{sm-Canham1970}%
  \BibitemOpen
  \bibfield  {author} {\bibinfo {author} {\bibfnamefont {P.~B.}\ \bibnamefont
  {Canham}},\ }\href {https://doi.org/10.1016/S0022-5193(70)80032-7} {\bibfield
   {journal} {\bibinfo  {journal} {J. Theor. Biol.}\ }\textbf {\bibinfo
  {volume} {26}},\ \bibinfo {pages} {61} (\bibinfo {year} {1970})}\BibitemShut
  {NoStop}%
\bibitem [{\citenamefont {Helfrich}(1973)}]{sm-Helfrich73}%
  \BibitemOpen
  \bibfield  {author} {\bibinfo {author} {\bibfnamefont {W.}~\bibnamefont
  {Helfrich}},\ }\href {\doibase 10.1515/znc-1973-11-1209} {\bibfield
  {journal} {\bibinfo  {journal} {Z. Naturforsch. C}\ }\textbf {\bibinfo
  {volume} {28}},\ \bibinfo {pages} {693} (\bibinfo {year} {1973})}\BibitemShut
  {NoStop}%
\bibitem [{\citenamefont {Evans}(1974)}]{sm-Evans74}%
  \BibitemOpen
  \bibfield  {author} {\bibinfo {author} {\bibfnamefont {E.~A.}\ \bibnamefont
  {Evans}},\ }\href {https://doi.org/10.1016/S0006-3495(74)85959-X} {\bibfield
  {journal} {\bibinfo  {journal} {Biophys. J.}\ }\textbf {\bibinfo {volume}
  {14}},\ \bibinfo {pages} {923} (\bibinfo {year} {1974})}\BibitemShut
  {NoStop}%
\bibitem [{\citenamefont {Monge}(1807)}]{sm-monge}%
  \BibitemOpen
  \bibfield  {author} {\bibinfo {author} {\bibfnamefont {G.}~\bibnamefont
  {Monge}},\ }\href@noop {} {\emph {\bibinfo {title} {Application de l'analyse
  \`{a} la g\'{e}om\'{e}trie}}}\ (\bibinfo  {publisher} {Bernard},\ \bibinfo
  {year} {1807})\BibitemShut {NoStop}%
\bibitem [{\citenamefont {Leal}(2007)}]{sm-Leal}%
  \BibitemOpen
  \bibfield  {author} {\bibinfo {author} {\bibfnamefont {L.~G.}\ \bibnamefont
  {Leal}},\ }\href {\doibase 10.1017/CBO9780511800245} {\emph {\bibinfo {title}
  {Advanced Transport Phenomena: Fluid Mechanics and Convective Transport
  Processes}}},\ Cambridge Series in Chemical Engineering\ (\bibinfo
  {publisher} {Cambridge University Press},\ \bibinfo {year}
  {2007})\BibitemShut {NoStop}%
\bibitem [{\citenamefont {Turlier}\ and\ \citenamefont
  {Betz}(2018)}]{sm-Turlier18}%
  \BibitemOpen
  \bibfield  {author} {\bibinfo {author} {\bibfnamefont {H.}~\bibnamefont
  {Turlier}}\ and\ \bibinfo {author} {\bibfnamefont {T.}~\bibnamefont {Betz}},\
  }\enquote {\bibinfo {title} {Fluctuations in active membranes},}\ in\ \href
  {\doibase 10.1007/978-3-030-00630-3_21} {\emph {\bibinfo {booktitle} {Physics
  of Biological Membranes}}},\ \bibinfo {editor} {edited by\ \bibinfo {editor}
  {\bibfnamefont {P.}~\bibnamefont {Bassereau}}\ and\ \bibinfo {editor}
  {\bibfnamefont {P.}~\bibnamefont {Sens}}}\ (\bibinfo  {publisher} {Springer
  International Publishing},\ \bibinfo {address} {Cham},\ \bibinfo {year}
  {2018})\ pp.\ \bibinfo {pages} {581--619}\BibitemShut {NoStop}%
\bibitem [{\citenamefont {Yan}\ and\ \citenamefont
  {Brady}(2015)}]{sm-yan2015force}%
  \BibitemOpen
  \bibfield  {author} {\bibinfo {author} {\bibfnamefont {W.}~\bibnamefont
  {Yan}}\ and\ \bibinfo {author} {\bibfnamefont {J.~F.}\ \bibnamefont
  {Brady}},\ }\href {\doibase 10.1017/jfm.2015.621} {\bibfield  {journal}
  {\bibinfo  {journal} {J. Fluid Mech.}\ }\textbf {\bibinfo {volume} {785}},\
  \bibinfo {pages} {R1} (\bibinfo {year} {2015})}\BibitemShut {NoStop}%
\bibitem [{\citenamefont {Nikola}\ \emph {et~al.}(2016)\citenamefont {Nikola},
  \citenamefont {Solon}, \citenamefont {Kafri}, \citenamefont {Kardar},
  \citenamefont {Tailleur},\ and\ \citenamefont
  {Voituriez}}]{sm-nikola2016active}%
  \BibitemOpen
  \bibfield  {author} {\bibinfo {author} {\bibfnamefont {N.}~\bibnamefont
  {Nikola}}, \bibinfo {author} {\bibfnamefont {A.~P.}\ \bibnamefont {Solon}},
  \bibinfo {author} {\bibfnamefont {Y.}~\bibnamefont {Kafri}}, \bibinfo
  {author} {\bibfnamefont {M.}~\bibnamefont {Kardar}}, \bibinfo {author}
  {\bibfnamefont {J.}~\bibnamefont {Tailleur}}, \ and\ \bibinfo {author}
  {\bibfnamefont {R.}~\bibnamefont {Voituriez}},\ }\href
  {10.1103/PhysRevLett.117.098001} {\bibfield  {journal} {\bibinfo  {journal}
  {Phys. Rev. Lett.}\ }\textbf {\bibinfo {volume} {117}},\ \bibinfo {pages}
  {098001} (\bibinfo {year} {2016})}\BibitemShut {NoStop}%
\end{thebibliography}
